\documentclass[twocolumn,titlepage,superscriptaddress,floatfix]{revtex4-1}

\usepackage{braket}
\usepackage{graphicx}
\usepackage{dcolumn}
\usepackage{bm}
\usepackage{times}
\usepackage{hyperref}

\usepackage{amssymb}

\usepackage{amsmath}
\usepackage{algorithm}
\usepackage{algpseudocode}
\usepackage{mathtools}

\usepackage[T1]{fontenc}
\usepackage[font={sf,small},labelfont=bf,figurename=Figure]{caption}
\usepackage[singlelinecheck=off,justification=raggedright]{subcaption}
\DeclareCaptionLabelSeparator{bar}{ | }
\usepackage{tikz}
\usetikzlibrary{positioning}
\definecolor {processblue}{cmyk}{0.96,0,0,0}
\definecolor {processyellow}{cmyk}{0,0,0.18,0}

\makeatletter
\def\BState{\State\hskip-\ALG@thistlm}
\makeatother

\makeatletter
\newcommand*\bigcdot{\mathpalette\bigcdot@{.6}}
\newcommand*\bigcdot@[2]{\mathbin{\vcenter{\hbox{\scalebox{#2}{$\m@th#1\bullet$}}}}}
\makeatother

\usepackage[compact]{titlesec}
\titleformat{\section}[hang]{\normalfont\sffamily\normalsize\bfseries}{}{0 pt}{}
\titleformat{\paragraph}[runin]{\normalfont\small\bfseries}{}{0 pt}{}
\titleformat{\subsection}[hang]{\normalfont\sffamily\small\bfseries}{}{0 pt}{}

\begin{document}

\title{Deterministic Bidirectional Communication and Remote Entanglement Generation Between Superconducting Quantum Processors}

\author{N. Leung}
\thanks{These authors contributed equally to this work.
\newline nelsonleung@uchicago.edu
\newline luy100@uchicago.edu}
\affiliation{The James Franck Institute and Department of Physics, University of Chicago, Chicago, Illinois 60637, USA}
\author{Y. Lu}
\thanks{These authors contributed equally to this work.
\newline nelsonleung@uchicago.edu
\newline luy100@uchicago.edu}
\affiliation{The James Franck Institute and Department of Physics, University of Chicago, Chicago, Illinois 60637, USA}
\author{S. Chakram}
\affiliation{The James Franck Institute and Department of Physics, University of Chicago, Chicago, Illinois 60637, USA}
\email{luy100@uchicago.edu}
\author{R. K. Naik}
\affiliation{The James Franck Institute and Department of Physics, University of Chicago, Chicago, Illinois 60637, USA}
\author{N. Earnest}
\affiliation{The James Franck Institute and Department of Physics, University of Chicago, Chicago, Illinois 60637, USA}
\author{R. Ma}
\affiliation{The James Franck Institute and Department of Physics, University of Chicago, Chicago, Illinois 60637, USA}
\author{K. Jacobs}
\affiliation{U.S. Army Research Laboratory, Computational and Information Sciences Directorate, Adelphi, Maryland 20783, USA}
\affiliation{Department of Physics, University of Massachusetts at Boston, Boston, MA 02125, USA}
\author{A. N. Cleland}
\affiliation{Institute for Molecular Engineering, University of Chicago, Chicago, Illinois 60637, USA}
\author{D. I. Schuster}
\email{david.schuster@uchicago.edu}
\affiliation{The James Franck Institute and Department of Physics, University of Chicago, Chicago, Illinois 60637, USA}

\pacs{Valid PACS appear here}
\maketitle

\section*{Abstract}

We propose and experimentally demonstrate a simple and efficient scheme for photonic communication between two remote superconducting modules. Each module consists of a random access quantum information processor with eight-qubit multimode memory and a single flux tunable transmon. The two processor chips are connected through a one-meter long coaxial cable that is coupled to a dedicated ``communication'' resonator on each chip. The two communication resonators hybridize with a mode of the cable to form a dark ``communication mode'' that is highly immune to decay in the coaxial cable. We modulate the transmon frequency via a parametric drive to generate sideband interactions between the transmon and the communication mode. We demonstrate bidirectional single-photon transfer with a success probability exceeding 60\%, and generate an entangled Bell pair with a fidelity of 79.3 $\pm$ 0.3\%.

\section*{Introduction}

A practical quantum computer requires a large number of qubits working in cooperation~\cite{Fowler2012SurfaceComputation}, a challenging task for any quantum hardware platform. For superconducting qubits, there is an ongoing effort to integrate increasing numbers of qubits on a single chip~\cite{Noroozian2012CrosstalkArrays, Wenner2011WirebondQubits, Chen2014FabricationCircuits, Vesterinen2014MitigatingQubits, Foxen2017QubitInterconnects,Dunsworth2018ADevices,Rosenberg20173DQubits}. A promising approach to scaling up  superconducting quantum computing  hardware is to adopt a modular architecture~\cite{Monroe2014Large-scaleInterconnects, Brecht2016MultilayerComputing, Chou2018DeterministicQubits} in which modules are connected together via communication channels to form a quantum network. This reduces the number of qubits required on a single chip, and allows greater flexibility in reconfiguring and extending the resulting information processing system. In such an architecture, each module is capable of performing universal operations on multiple-bits, and neighboring modules are connected through photonic channels, allowing communication and entanglement generation between remote modules.


Remote entanglement between superconducting qubits has been realized probabilistically~\cite{Roch2014ObservationQubits, Narla2016RobustQubits, Dickel2018Chip-to-chipFields}. Conversely, realizing deterministic photonic communication requires releasing a single photon from one qubit and catching it with the remote qubit. In the long-distance limit, the photon emission and absorption are from a continuum density of states. In this limit, static coupling limits the maximum transfer fidelity to only 54\% \cite{Stobinska2009PerfectSpace,Wang2011EfficientMode}. This limit is exceeded by dynamically tailoring the emission and absorption profiles ~\cite{Yin2013CatchStates, Srinivasan2014Time-reversalTransfer, Pechal2014Microwave-ControlledElectrodynamics, Wenner2014CatchingEfficiency}. These capabilities are presently being used to perform photonic communication between superconducting qubits connected by a transmission line within a cryostat~\cite{Axline2017On-demandMemories, Campagne-Ibarcq2017DeterministicTransitions, Dickel2017Chip-to-chipFields, Kurpiers2017DeterministicPhotons}. In these experiments, the use of a circulator enables the finite-length transmission line to be modeled as a long line with a continuum density of states, at the cost of added transmission loss.

Here, we establish bidirectional photonic communication between two
multi-qubit superconducting quantum processors through a multimodal communication channel. Rather than inserting a circulator, the multimode nature of the finite length transmission line is made manifest and exploited ~\cite{Jacobs2016FastNetworks}. For intra-cryostat communication, the required connection coaxial cable length of 1 m or less results in a free spectral range on the order of hundreds of MHz. In this setting, the resonances of the coaxial cable form hybridized normal modes with on-chip communication resonators, and photons are transferred coherently through the discrete modes of the channel in contrast to emission/absorption through a continuum. We use parametric flux modulation of the qubit frequency to generate resonant sideband interactions between the qubit and the communication channel ~\cite{Beaudoin2012First-orderModulation, Strand2013, Sirois2015Coherent-stateConversion, McKay2016UniversalBus}. This approach avoids the loss
due to the circulator that significantly limits the communication fidelity, and enables bidirectional quantum communication.

\section*{Results}

\begin{figure*}
\centering 

\includegraphics[width=2.0\columnwidth]{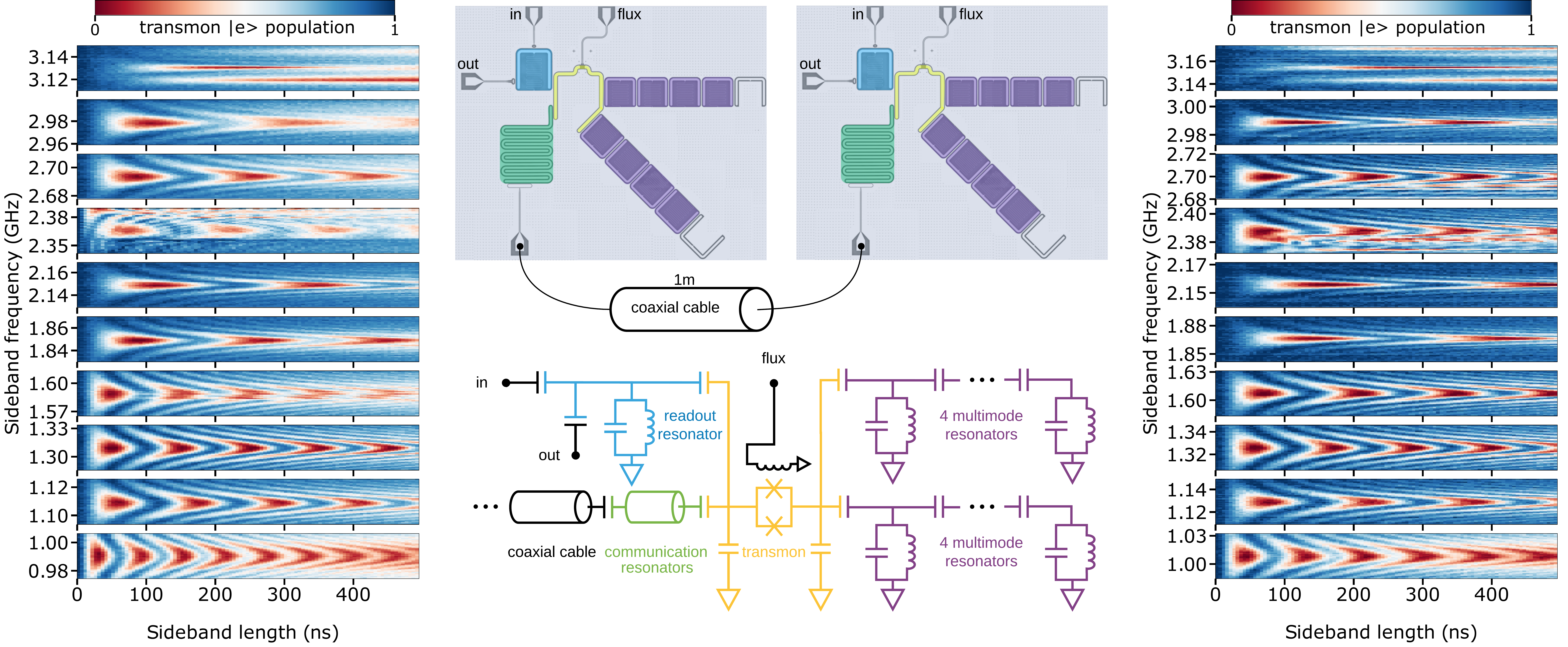}

  \caption{{\textbf{Device schematic and stimulated vacuum Rabi oscillations.}} Each chip consists of a frequency-tunable transmon and two chains of four identically designed, lumped-element resonators. In addition, a resonator is included for readout, and a second resonator is coupled to the coaxial cable ($\sim$ 1 m) that provides the communication link between the chips. The simple circuit diagram shows the circuit model of a each module. We induce resonant interactions between the transmon and an individual mode by modulating the transmon frequency via its flux bias at the frequency difference (detuning) between the mode and the transmon. The chevron patterns indicate parametrically induced resonant oscillations with each of the modes. These patterns are generated by sweeping the length of the flux modulation pulse at each frequency and measuring the excited state population of the transmon after the pulse ends. The sideband modulation of the frequency-tunable transmon can access resonators in each chip by targeting their corresponding frequency detunings. On each chip, the lowest frequency corresponds to the readout resonator, and the highest frequency corresponds to the communication resonator. The eight memory mode frequencies are intermediate to these and spaced by $\approx$ 200\,MHz.}
\label{circuit}
\end{figure*}

\subsection*{Network of two multimode modules}
We extend the random access quantum processor module presented in Ref.~\cite{Naik2017RandomElectrodynamics} to allow photonic communication between two remote modules, thereby realizing a two-node quantum network. Each processor consists of an eight-qubit multimode memory comprised of two chains of four identical and strongly coupled superconducting resonators, a single flux-tunable transmon, and two additional resonators~\cite{McKay2015High-ContrastQED}. The first of these resonators is used for readout, and the second is coupled to the coaxial cable to enable the inter-module communication. The transmon can resonantly couple to all the resonators (readout, multimode and communication) through parametric flux modulation to realize intra-module gate operations and inter-module photonic communications. Figure~\ref{circuit} shows a schematic of our two modules. The readout resonators have the lowest frequencies [module 1: 5.7463\,GHz; module 2: 5.7405\,GHz], the communication resonators have the highest frequencies [$\approx$ 7.88\,GHz, see the appendix for detailed analysis of parameters] , and the eight memory mode frequencies are in the range 5.8\,GHz - 7.7\,GHz, spaced by $\approx$ 200\,MHz. For the circuit design, we arranged the multimode resonators to be spatially separated from the readout and communication resonators by placing the high impedance transmon in-between, preventing Purcell loss of the multimode resonators through the low Q readout and communication resonators \cite{Houck2008ControllingQubit, Reed2010FastQubit, Gambetta2011SuperconductingCoupling}. We operate the transmons at the static frequency of [1: 4.7685\,GHz; 2: 4.7420\,GHz] with an anharmonicity of [1: 109.8\,MHz; 2: 109.9\,MHz]. 

We induce resonant interactions between the transmon and an individual mode by modulating the transmon frequency via its flux bias. The modulation creates sidebands of the transmon excited state, detuned from the original resonance by the frequency of the applied flux tone. When one of these sidebands is resonant with a mode, the system experiences stimulated vacuum Rabi oscillations~\cite{Naik2017RandomElectrodynamics}. This process is similar to resonant vacuum Rabi oscillations~\cite{Rempe1987ObservationMaser}, but occur at a rate that is controlled by the modulation amplitude~\cite{Strand2013,Beaudoin2012First-orderModulation}. To illustrate the application of parametric control, we employ the following experimental sequence. First, the transmon is excited via its charge bias. Subsequently, we modulate the flux bias to create sidebands of the transmon excited state at the modulation frequency. This is repeated for different flux pulse durations and frequencies, with the population of the transmon excited state measured at the end of each sequence. When the frequency matches the detuning between the transmon and a given eigenmode, we observe full-contrast stimulated vacuum Rabi oscillations. Figure \ref{circuit} shows that the transmon can selectively interact with each of the eigenmodes by choosing the appropriate modulation frequency. As previously demonstrated, this sideband interaction and rotations of the transmon are sufficient for universal operations on each set of multimode resonators \cite{Naik2017RandomElectrodynamics}. Similarly, the photon transfer process between two remote qubits is initiated by switching on the sideband interactions targeting the communication resonator on each chip. As the bare frequencies of the transmon and the communication resonator are far detuned ($\Delta \approx$ 3\,GHz, $g \approx$ 50\,MHz), the sideband coupling scheme for photonic communication achieves a high on/off ratio.

\begin{figure}
\centering 
\includegraphics[width=1.0\columnwidth]{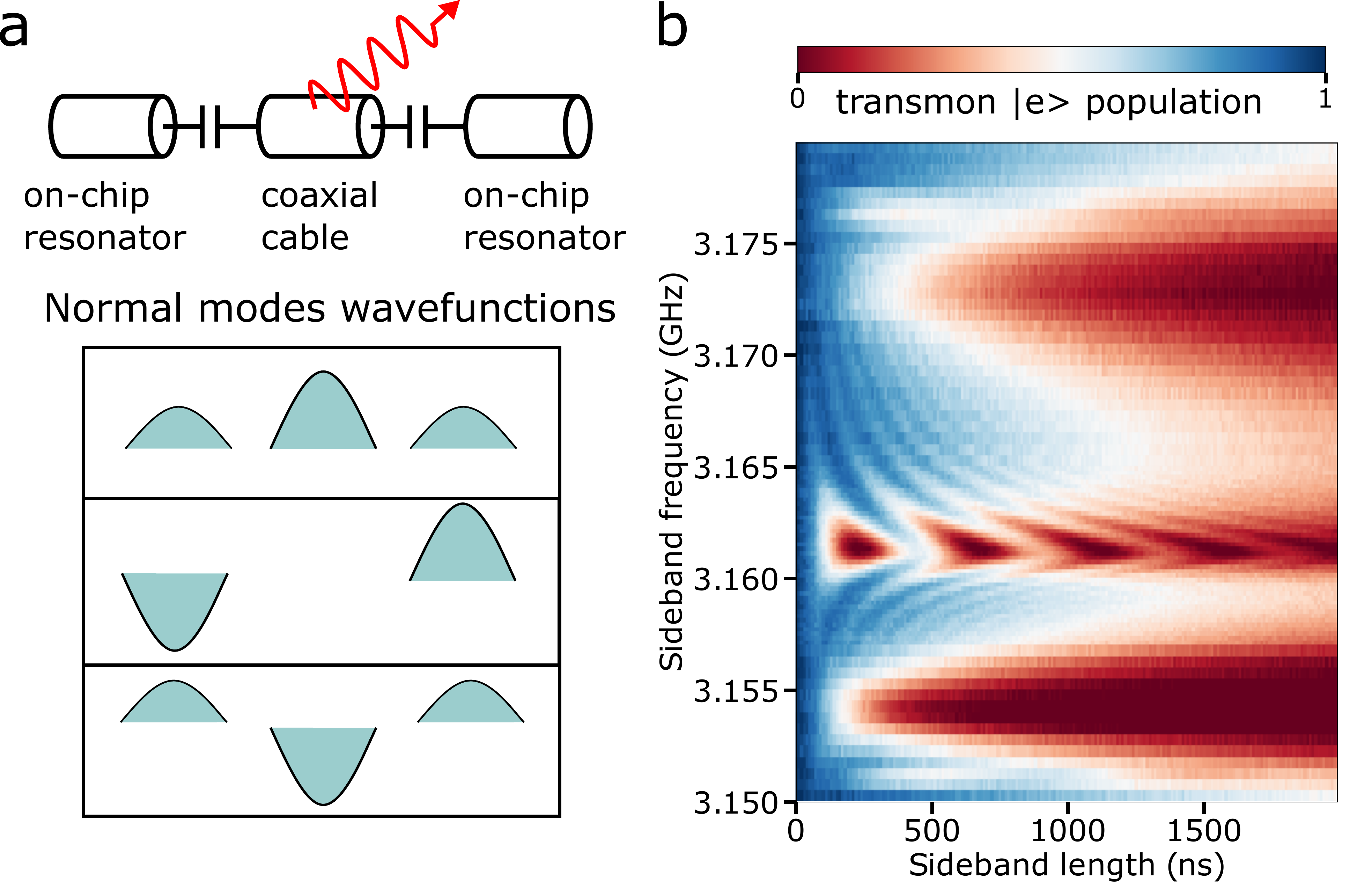}
  \caption{{\textbf{Hybridized normal modes. }}  \textbf{a.} The schematic showing the wavefunctions of the coupled system involving the communication resonators and the coaxial cable. The three degenerate modes hybridize and form three normal modes with distinct signatures. The center normal mode has minimal participation in the lossy cable mode and has high quality factor. \textbf{b.} Stimulated vacuum Rabi oscillations around the communication modes. The near-degeneracy of the coaxial cable with the two communication resonators give rise to this almost equally-spaced three-mode structure. Being the two bright modes that include the lossy cable mode, and the dark ``communication'' mode of the two resonators. The latter couples more strongly to both qubits, and has a lifetime that is ideally only limited by the internal quality factors of the communication resonators. By fitting the simulation to experimental data, we found that the coaxial cable has a slightly higher frequency than the on-chip communication resonators [see Appendix]. }
\label{comm_coherence}
\end{figure}






\subsection*{Multimode communication channel}

The two communication resonators are designed to have identical frequencies. They are chosen to be coplanar waveguide resonators with a large center pin and gap width to make the frequency insensitive to fabrication variations~\cite{Underwood2012Low-disorderPhotons}.  These resonators are coupled via the one-meter long coaxial cable, where the cable can be thought of as a multimode resonator with a free spectral range of around 200\,MHz. The coupling strength between the cable and the communication resonators is $g_c\approx 7~\mbox{MHz}$. The cable mode that we use for communication has a frequency that is within $g_c$ of the frequencies of the communication modes. Since the free spectral range of the coaxial cable is an order of magnitude larger than $g_c$, we consider the cable as a single mode nearly resonant with the communication resonators. The cable and the communication resonators thus together produce three hybridized normal modes which are depicted in Figure~\ref{comm_coherence}. The near-degeneracy of the coaxial cable with the two communication resonators give rise to this almost equally-spaced three-mode structure, which can be seen from the three stimulated vacuum Rabi chevrons in Fig.~\ref{comm_coherence} b. The center normal mode used for communication ideally has no participation in the cable mode, and as a result, its loss rate is limited by the internal quality factors of the communication resonators and small Purcell losses from neighboring cable modes. In comparison to the neighboring modes, the center normal mode couples more strongly to both qubits due to higher wavefunction participation at the communication resonators. Thus, this communication mode has both the advantages of high quality factor and high coupling rate. For any practical device, the center normal mode does have a non-zero participation in the lossy coaxial cable due to a frequency mismatch between the two on-chip communication resonators. From the measurements of previous individual test chips, the detuning between these two resonators is expected to be less than 3\,MHz ($< g_c$), an assumption that is validated by the simulation shown in the appendix, resulting in a less than 5\% of cable mode participation in the communication mode.

The coherence time of the communication mode can be characterized using protocols analogous to those for
the transmon; the qubit pulses are merely sandwiched between a pair of transmon-mode iSWAP pulses to transfer the quantum state
between the transmon and the mode~\cite{Naik2017RandomElectrodynamics}. We find $T_1 = 550$\,ns  and $T_2^* = 1$\,$\mu$s, corresponding to a quality factor of about 4000. This quality factor is reasonably high, considering that it involves losses from the long lossy cable, wirebonds, solder of the SMA connector, and the copper leads of the sample holder. The two neighboring normal modes have much lower coherence times due to the higher participation of the lossy cable mode. From fitting to fig.~\ref{comm_coherence}b we estimate an upper bound of $T_1$ for these modes to be $\sim$ 200\,ns.

\begin{figure}[]
  \centering \includegraphics[width=\columnwidth]{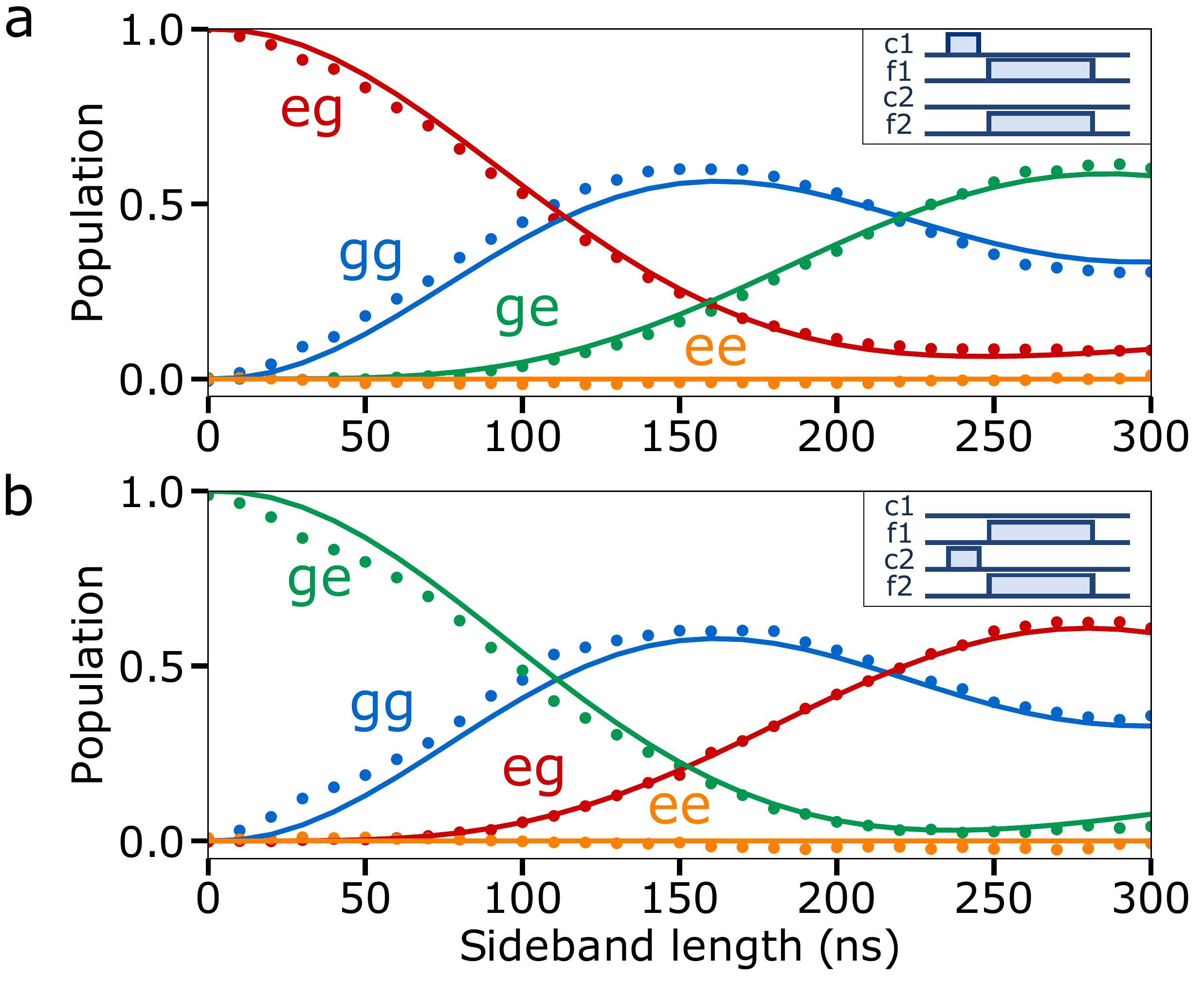}
  \caption{{\textbf{Bidirectional excitation transfer. }} The inset at top right shows the pulse sequence used to implement excitation transfer. The labels $c1$, $c2$ denote the charge drives on qubits 1 and 2, respectively, and $f1$, $f2$ the respective flux drives. We first apply a $\pi$ pulse to excite one of the qubits, then simultaneously switch on the sideband flux pulse to drive the transfer process. Using the same sideband sequence, but instead applying the $\pi$ pulse to the other qubit, we can send a single photon in the opposite direction. The transfer fidelity is limited by qubit dephasing and photon decay in the communication mode. Described in the following, the transfer process in different directions have slightly different loss mechanisms. \textbf{a.} Excitation transfer from qubit 1 to qubit 2. Notice that in this transfer process the sender qubit is not able to fully receive its excitation (population of $|eg\rangle$ does not reach zero). As confirmed by the master equation simulation, this is due to the dephasing of qubit 1. The remaining errors arise from communication cavity loss and dephasing of qubit 2, which is less than that of qubit 1. \textbf{b.} Excitation transfer from qubit 2 to qubit 1. In this process, while qubit 2 releases most of its excitations (population of $|ge\rangle$ comes close to zero), the dephasing of qubit 1 prevents it from capturing all the excitations in the communication mode, resulting in a slightly higher final population in $|gg\rangle$. The resulting fidelities for the transfer in the two directions are similar: $\{P_{|ge\rangle}, P_{|eg\rangle}  \} \approx $ 61\%, confirming the results from our numerical simulation. }
  \label{two_way}
\end{figure}

\subsection*{Bidirectional communication}

To demonstrate photonic communication between the two chips, we send a single photon from one chip to the other. First, we excite the sender qubit, then we switch on sideband interactions simultaneously on both qubits, targeting the communication channel. We send a photon in the reverse direction using the same sideband sequence but instead exciting the other qubit, thus demonstrating bidirectional photon transfer. Figure~\ref{two_way} shows the transmon population plotted as a function of the sideband pulse length. The master equation simulation results (solid lines) are shown along with the experimental data (dots). We are able to obtain photon transfer with a success rate of $\{P_{|ge\rangle}, P_{|eg\rangle}  \} \approx $ 61\%. We use simultaneous square pulses for the time-envelopes of the sideband interactions. From the simulations detailed in the appendix, we found that square pulses gave superior performance for our current circuit parameters. Note that the achieved  transfer fidelity exceeds 54\%, the maximum fidelity for absorbing a naturally shaped emission into a continuum~\cite{Stobinska2009PerfectSpace,Wang2011EfficientMode}. This demonstrates a qualitative difference in transferring via a multimode cable compared to that of releasing and catching flying photonic qubits through a continuum. 

The transfer fidelity is limited by qubit dephasing and photon decay in the communication mode. Qubit 1 has a higher dephasing rate ($T_2^* \approx$ 700\,ns) than qubit 2 ($T_2^* \approx$ 1.4\,$\mu$s). The dephasing rate of qubit 1 is comparable to the sideband coupling rate, with the result that this qubit is not able to fully release its excitation during the transfer process. Conversely, for transfer in the other direction qubit 1 is not able to receive all of the excitations. This transfer infidelity can be largely mitigated by using a fixed-frequency qubit less susceptible to the flux noise, with its coupling strength to the communication mode parametrically controlled via a tunable coupler circuit \cite{Allman2014TunableResonator,Chen2014QubitCoupling,Sirois2015Coherent-stateConversionb,McKay2016UniversalBusb,Lu2017UniversalQubitb}. The remaining loss of transfer fidelity comes from the loss in the communication mode. From our numerical simulations detailed in the appendix, we estimate that the overall photon loss in both the qubits and the communication mode contribute to an infidelity of 24\%, while the dephasing error of the two qubits accounts for an infidelity of 15\%. The sideband coupling rate of the transmon is limited by the range over which its frequency can be parametrically tuned, resulting in a maximum effective sideband coupling to the communication resonator of $\approx$ 2\,MHz. With improved qubit coherence time, our simulation shows that more sophisticated transfer protocols such as STIRAP~\cite{Halfmann1998CoherentSO2, Vasilev2009OptimumPassage} can be employed to boost transfer efficiency.

\begin{figure}[]
  \centering \includegraphics[width=\columnwidth]{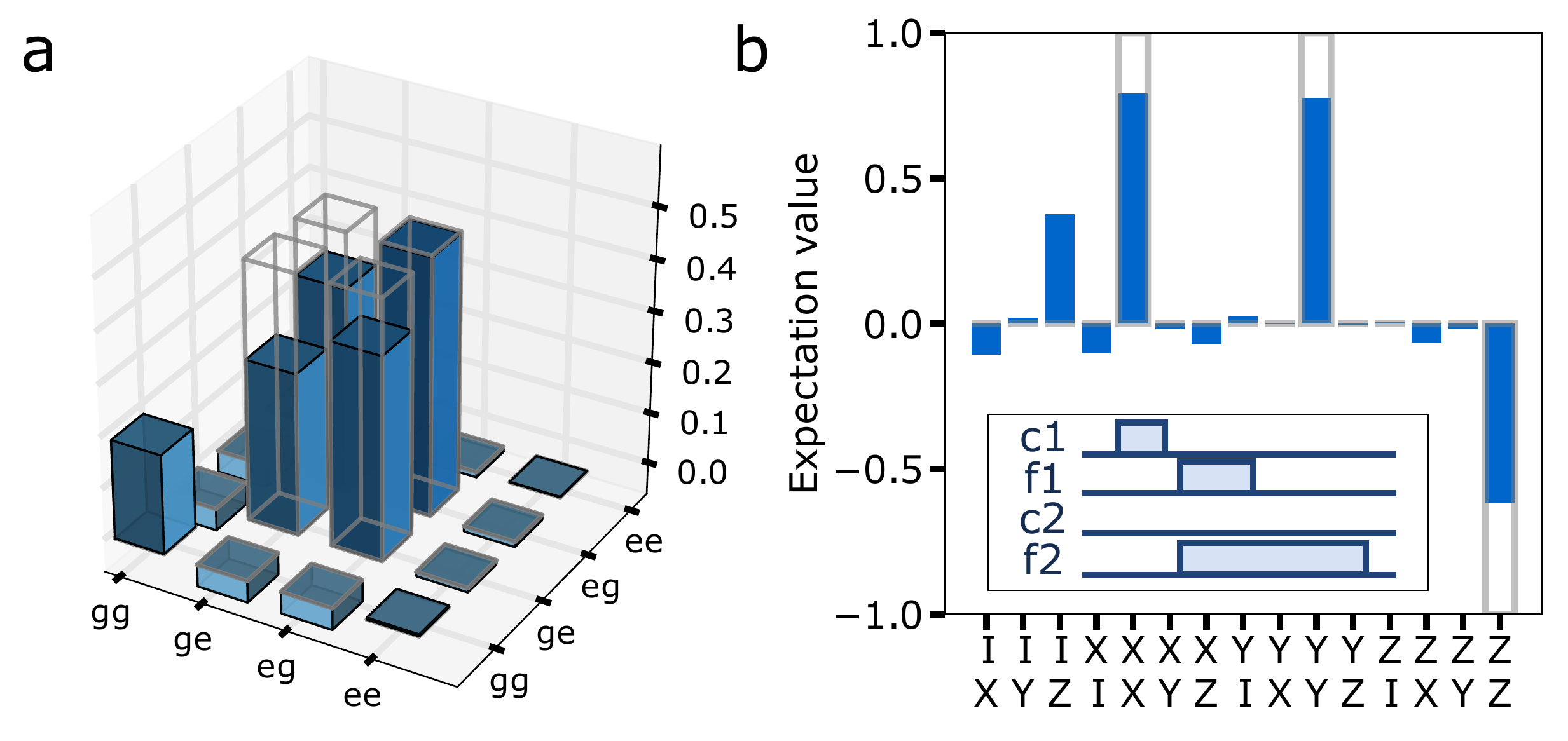}
  \caption{{\textbf{Bell pair creation.}} \textbf{a.} Real component of the density matrix. \textbf{b.} Expectation values of two-qubit Pauli operators. We create a Bell state between two remote qubits, one on each module. This is achieved by first applying the $\sqrt{i\textnormal{SWAP}}$ gate between the excited qubit 1 and the communication mode, which is implemented by a sideband modulation pulse to qubit 1 to perform a $\pi/2$ rotation. A similar pulse, this time a $\pi$ rotation, applied to the second qubit performs an $i\textnormal{SWAP}$ that transfers the entanglement from the communimation mode to the second qubit.  As shown in the inset, we implement the two pulses simultaneously to reduce decoherence. We obtain the resulting Bell state with fidelity $\langle \Psi^+| \rho_{\textnormal{exp}}|\Psi^+\rangle = $ 79.3 $\pm$ 0.3 \%. This fidelity is significantly higher than the transfer probability that we achieve for a single excitation. This can be understood by noting that to create the Bell state the operation involving qubit 1 takes approximately half the time, the excitation does not need to be fully transferred from qubit 1, and there is less excitation in the cable over the duration of the protocol.  }\label{Bell}
\end{figure}

\subsection*{Bell state entanglement}

We now entangle two qubits by creating a Bell state between the transmons on the respective chips~\cite{Bell1964OnParadox}. We can create such a state by first applying the $\sqrt{i\textnormal{SWAP}}$ gate between the excited qubit 1 and the communication mode, which generates the Bell state $(|g1\rangle + |e0\rangle)/\sqrt{2}$ between them. We implement the $\sqrt{i\textnormal{SWAP}}$ by applying a sideband modulation pulse to qubit 1 to perform a $\pi/2$ rotation. Subsequently, we transfer the state of the communication mode to qubit 2 through the $i\textnormal{SWAP}$ gate by applying a sideband modulation pulse to the latter to perform a $\pi$ rotation. Ideally this sequence prepares the Bell state $|\Psi^+\rangle =  (|ge\rangle + |eg\rangle)/\sqrt{2}$ shared between the two remote qubits. To minimize decoherence the sender and receiver pulses can be applied simultaneously, so long as the lengths and amplitudes of the pulses are adjusted appropriately. Choosing qubit 1 as the sender and using square pulses, we found --- both in our simulation and in the experiment --- that maximal fidelity was obtained by setting both pulses at the same coupling rate and the length of the receiver pulse to be slightly longer than twice that of the sender. The resulting Bell state has a fidelity of $\langle \Psi^+| \rho_{\textnormal{exp}}|\Psi^+\rangle = $ 79.3 $\pm$ 0.3\%. We obtained the density matrix $\rho_{\textnormal{exp}}$ using quantum state tomography with an over-complete set of measurements complemented with the maximum likelihood method~\cite{James2001MeasurementQubits}. It can be inferred from the data that the fidelity is almost equally limited by photon decay in the cable and the qubit dephasing errors. We also note that the Bell state fidelity is significantly higher than the success probability we achieved for  photon transfer. Likely explanations for this is that qubit 1 is actively involved in the process for only half the duration of the protocol and there is less excitation in the cable over the duration of the protocol. The process is thus less sensitive to the dephasing noise in the qubit and decay loss in the cable.

\section*{Conclusion}

We have built upon the random access quantum information processor previously presented in Ref.~\cite{Naik2017RandomElectrodynamics}, so as to realize photonic communication between two remote modules, a first step in realizing a modular network. The sideband modulation of the transmon qubit in each module can be applied to implement local operations on the multimode resonators and to perform photon transfer between the two modules. The multimode characteristic of the communication channel (a coaxial cable) is enabled by the absence of a circulator. This mode structure results in normal modes that are superpositions of a mode of the inter-module communication cable and the on-chip resonators. One of the these normal modes is ``dark'' to the coaxial cable mode, thus avoiding much of the cable loss and allowing for
high fidelity photon transfer. We characterized our system by performing single photon transfer with 61\% fidelity and Bell-state preparation with 79.3\% fidelity. These fidelities can be increased by improving the qubit coherence time and the strength of the coupling to the communication channel. Future work will include implementing more sophisticated photon transfer protocols (e.g. STIRAP), applying heralding protocols to protect against photon transmission error, implementing local gates on memory modes in conjunction with photonic communication to facilitate large-scale computation, and integrating the present architecture with high-quality-factor 3D superconducting cavities \cite{Reagor2013ReachingCavities}.




\section*{Data availability}

Data available on request from authors.

\bibliography{Mendeley_communication_NL_new}

\begin{thebibliography}{10}
\expandafter\ifx\csname url\endcsname\relax
  \def\url#1{\texttt{#1}}\fi
\expandafter\ifx\csname urlprefix\endcsname\relax\def\urlprefix{URL }\fi
\providecommand{\bibinfo}[2]{#2}
\providecommand{\eprint}[2][]{\url{#2}}

\bibitem{Fowler2012SurfaceComputation}
\bibinfo{author}{Fowler, A.~G.}, \bibinfo{author}{Mariantoni, M.},
  \bibinfo{author}{Martinis, J.~M.} \& \bibinfo{author}{Cleland, A.~N.}
\newblock \bibinfo{title}{{Surface codes: Towards practical large-scale quantum
  computation}}.
\newblock \emph{\bibinfo{journal}{Physical Review A}}
  \textbf{\bibinfo{volume}{86}}, \bibinfo{pages}{032324}
  (\bibinfo{year}{2012}).

\bibitem{Noroozian2012CrosstalkArrays}
\bibinfo{author}{Noroozian, O.}, \bibinfo{author}{Day, P.~K.},
  \bibinfo{author}{Eom, B.~H.}, \bibinfo{author}{Leduc, H.~G.} \&
  \bibinfo{author}{Zmuidzinas, J.}
\newblock \bibinfo{title}{{Crosstalk Reduction for Superconducting Microwave
  Resonator Arrays}}.
\newblock \emph{\bibinfo{journal}{IEEE Transactions on Microwave Theory and
  Techniques}} \textbf{\bibinfo{volume}{60}}, \bibinfo{pages}{1235--1243}
  (\bibinfo{year}{2012}).

\bibitem{Wenner2011WirebondQubits}
\bibinfo{author}{Wenner, J.} \emph{et~al.}
\newblock \bibinfo{title}{{Wirebond crosstalk and cavity modes in large chip
  mounts for superconducting qubits}}.
\newblock \emph{\bibinfo{journal}{Superconductor Science and Technology}}
  \textbf{\bibinfo{volume}{24}}, \bibinfo{pages}{065001}
  (\bibinfo{year}{2011}).

\bibitem{Chen2014FabricationCircuits}
\bibinfo{author}{Chen, Z.} \emph{et~al.}
\newblock \bibinfo{title}{{Fabrication and characterization of aluminum
  airbridges for superconducting microwave circuits}}.
\newblock \emph{\bibinfo{journal}{Applied Physics Letters}}
  \textbf{\bibinfo{volume}{104}}, \bibinfo{pages}{052602}
  (\bibinfo{year}{2014}).

\bibitem{Vesterinen2014MitigatingQubits}
\bibinfo{author}{Vesterinen, V.}, \bibinfo{author}{Saira, O.~P.},
  \bibinfo{author}{Bruno, A.} \& \bibinfo{author}{DiCarlo, L.}
\newblock \bibinfo{title}{{Mitigating information leakage in a crowded spectrum
  of weakly anharmonic qubits}}  (\bibinfo{year}{2014}).
\newblock \eprint{\textit{arXiv:1405.0450}}.

\bibitem{Foxen2017QubitInterconnects}
\bibinfo{author}{Foxen, B.} \emph{et~al.}
\newblock \bibinfo{title}{{Qubit compatible superconducting interconnects}}
  (\bibinfo{year}{2017}).
\newblock \eprint{\textit{arXiv:1708.04270}}.

\bibitem{Dunsworth2018ADevices}
\bibinfo{author}{Dunsworth, A.} \emph{et~al.}
\newblock \bibinfo{title}{{A method for building low loss multi-layer wiring
  for superconducting microwave devices}}.
\newblock \emph{\bibinfo{journal}{Applied Physics Letters}}
  \textbf{\bibinfo{volume}{112}}, \bibinfo{pages}{063502}
  (\bibinfo{year}{2018}).

\bibitem{Rosenberg20173DQubits}
\bibinfo{author}{Rosenberg, D.} \emph{et~al.}
\newblock \bibinfo{title}{{3D integrated superconducting qubits}}.
\newblock \emph{\bibinfo{journal}{npj Quantum Information}}
  \textbf{\bibinfo{volume}{3}}, \bibinfo{pages}{42} (\bibinfo{year}{2017}).

\bibitem{Monroe2014Large-scaleInterconnects}
\bibinfo{author}{Monroe, C.} \emph{et~al.}
\newblock \bibinfo{title}{{Large-scale modular quantum-computer architecture
  with atomic memory and photonic interconnects}}.
\newblock \emph{\bibinfo{journal}{Physical Review A}}
  \textbf{\bibinfo{volume}{89}}, \bibinfo{pages}{022317}
  (\bibinfo{year}{2014}).

\bibitem{Brecht2016MultilayerComputing}
\bibinfo{author}{Brecht, T.} \emph{et~al.}
\newblock \bibinfo{title}{{Multilayer microwave integrated quantum circuits for
  scalable quantum computing}}.
\newblock \emph{\bibinfo{journal}{npj Quantum Information}}
  \textbf{\bibinfo{volume}{2}}, \bibinfo{pages}{16002} (\bibinfo{year}{2016}).

\bibitem{Chou2018DeterministicQubits}
\bibinfo{author}{Chou, K.~S.} \emph{et~al.}
\newblock \bibinfo{title}{{Deterministic teleportation of a quantum gate
  between two logical qubits}}  (\bibinfo{year}{2018}).
\newblock \eprint{\textit{arXiv:1801.05283}}.

\bibitem{Roch2014ObservationQubits}
\bibinfo{author}{Roch, N.} \emph{et~al.}
\newblock \bibinfo{title}{{Observation of Measurement-Induced Entanglement and
  Quantum Trajectories of Remote Superconducting Qubits}}.
\newblock \emph{\bibinfo{journal}{Physical Review Letters}}
  \textbf{\bibinfo{volume}{112}}, \bibinfo{pages}{170501}
  (\bibinfo{year}{2014}).

\bibitem{Narla2016RobustQubits}
\bibinfo{author}{Narla, A.} \emph{et~al.}
\newblock \bibinfo{title}{{Robust Concurrent Remote Entanglement Between Two
  Superconducting Qubits}}.
\newblock \emph{\bibinfo{journal}{Physical Review X}}
  \textbf{\bibinfo{volume}{6}}, \bibinfo{pages}{031036} (\bibinfo{year}{2016}).

\bibitem{Dickel2018Chip-to-chipFields}
\bibinfo{author}{Dickel, C.} \emph{et~al.}
\newblock \bibinfo{title}{{Chip-to-chip entanglement of transmon qubits using
  engineered measurement fields}}.
\newblock \emph{\bibinfo{journal}{Physical Review B}}
  \textbf{\bibinfo{volume}{97}}, \bibinfo{pages}{064508}
  (\bibinfo{year}{2018}).

\bibitem{Stobinska2009PerfectSpace}
\bibinfo{author}{Stobi{\'{n}}ska, M.}, \bibinfo{author}{Alber, G.} \&
  \bibinfo{author}{Leuchs, G.}
\newblock \bibinfo{title}{{Perfect excitation of a matter qubit by a single
  photon in free space}}.
\newblock \emph{\bibinfo{journal}{EPL (Europhysics Letters)}}
  \textbf{\bibinfo{volume}{86}}, \bibinfo{pages}{14007} (\bibinfo{year}{2009}).

\bibitem{Wang2011EfficientMode}
\bibinfo{author}{Wang, Y.}, \bibinfo{author}{Min{\'{a}}{\v{r}}, J.},
  \bibinfo{author}{Sheridan, L.} \& \bibinfo{author}{Scarani, V.}
\newblock \bibinfo{title}{{Efficient excitation of a two-level atom by a single
  photon in a propagating mode}}.
\newblock \emph{\bibinfo{journal}{Physical Review A}}
  \textbf{\bibinfo{volume}{83}}, \bibinfo{pages}{063842}
  (\bibinfo{year}{2011}).

\bibitem{Yin2013CatchStates}
\bibinfo{author}{Yin, Y.} \emph{et~al.}
\newblock \bibinfo{title}{{Catch and Release of Microwave Photon States}}.
\newblock \emph{\bibinfo{journal}{Physical Review Letters}}
  \textbf{\bibinfo{volume}{110}}, \bibinfo{pages}{107001}
  (\bibinfo{year}{2013}).

\bibitem{Srinivasan2014Time-reversalTransfer}
\bibinfo{author}{Srinivasan, S.~J.} \emph{et~al.}
\newblock \bibinfo{title}{{Time-reversal symmetrization of spontaneous emission
  for quantum state transfer}}.
\newblock \emph{\bibinfo{journal}{Physical Review A}}
  \textbf{\bibinfo{volume}{89}}, \bibinfo{pages}{033857}
  (\bibinfo{year}{2014}).

\bibitem{Pechal2014Microwave-ControlledElectrodynamics}
\bibinfo{author}{Pechal, M.} \emph{et~al.}
\newblock \bibinfo{title}{{Microwave-Controlled Generation of Shaped Single
  Photons in Circuit Quantum Electrodynamics}}.
\newblock \emph{\bibinfo{journal}{Physical Review X}}
  \textbf{\bibinfo{volume}{4}}, \bibinfo{pages}{041010} (\bibinfo{year}{2014}).

\bibitem{Wenner2014CatchingEfficiency}
\bibinfo{author}{Wenner, J.} \emph{et~al.}
\newblock \bibinfo{title}{{Catching Time-Reversed Microwave Coherent State
  Photons with 99.4{\%} Absorption Efficiency}}.
\newblock \emph{\bibinfo{journal}{Physical Review Letters}}
  \textbf{\bibinfo{volume}{112}}, \bibinfo{pages}{210501}
  (\bibinfo{year}{2014}).

\bibitem{Axline2017On-demandMemories}
\bibinfo{author}{Axline, C.} \emph{et~al.}
\newblock \bibinfo{title}{{On-demand quantum state transfer and entanglement
  between remote microwave cavity memories}}  (\bibinfo{year}{2017}).
\newblock \eprint{\textit{arXiv:1712.05832}}.

\bibitem{Campagne-Ibarcq2017DeterministicTransitions}
\bibinfo{author}{Campagne-Ibarcq, P.} \emph{et~al.}
\newblock \bibinfo{title}{{Deterministic remote entanglement of superconducting
  circuits through microwave two-photon transitions}}  (\bibinfo{year}{2017}).
\newblock \eprint{\textit{arXiv:1712.05854}}.

\bibitem{Dickel2017Chip-to-chipFields}
\bibinfo{author}{Dickel, C.} \emph{et~al.}
\newblock \bibinfo{title}{{Chip-to-chip entanglement of transmon qubits using
  engineered measurement fields}}  (\bibinfo{year}{2017}).
\newblock \eprint{\textit{arXiv:1712.06141}}.

\bibitem{Kurpiers2017DeterministicPhotons}
\bibinfo{author}{Kurpiers, P.} \emph{et~al.}
\newblock \bibinfo{title}{{Deterministic Quantum State Transfer and Generation
  of Remote Entanglement using Microwave Photons}}  (\bibinfo{year}{2017}).
\newblock \eprint{\textit{arXiv:1712.08593}}.

\bibitem{Jacobs2016FastNetworks}
\bibinfo{author}{Jacobs, K.} \emph{et~al.}
\newblock \bibinfo{title}{{Fast quantum communication in linear networks}}.
\newblock \emph{\bibinfo{journal}{EPL (Europhysics Letters)}}
  \textbf{\bibinfo{volume}{114}}, \bibinfo{pages}{40007}
  (\bibinfo{year}{2016}).

\bibitem{Beaudoin2012First-orderModulation}
\bibinfo{author}{Beaudoin, F.}, \bibinfo{author}{da~Silva, M.~P.},
  \bibinfo{author}{Dutton, Z.} \& \bibinfo{author}{Blais, A.}
\newblock \bibinfo{title}{{First-order sidebands in circuit QED using qubit
  frequency modulation}}.
\newblock \emph{\bibinfo{journal}{Physical Review A}}
  \textbf{\bibinfo{volume}{86}}, \bibinfo{pages}{022305}
  (\bibinfo{year}{2012}).

\bibitem{Strand2013}
\bibinfo{author}{Strand, J.~D.} \emph{et~al.}
\newblock \bibinfo{title}{First-order sideband transitions with flux-driven
  asymmetric transmon qubits}.
\newblock \emph{\bibinfo{journal}{Phys. Rev. B}} \textbf{\bibinfo{volume}{87}},
  \bibinfo{pages}{220505} (\bibinfo{year}{2013}).

\bibitem{Sirois2015Coherent-stateConversion}
\bibinfo{author}{Sirois, A.~J.} \emph{et~al.}
\newblock \bibinfo{title}{{Coherent-state storage and retrieval between
  superconducting cavities using parametric frequency conversion}}.
\newblock \emph{\bibinfo{journal}{Applied Physics Letters}}
  \textbf{\bibinfo{volume}{106}}, \bibinfo{pages}{172603}
  (\bibinfo{year}{2015}).

\bibitem{McKay2016UniversalBus}
\bibinfo{author}{McKay, D.~C.} \emph{et~al.}
\newblock \bibinfo{title}{{Universal Gate for Fixed-Frequency Qubits via a
  Tunable Bus}}.
\newblock \emph{\bibinfo{journal}{Physical Review Applied}}
  \textbf{\bibinfo{volume}{6}}, \bibinfo{pages}{064007} (\bibinfo{year}{2016}).

\bibitem{Naik2017RandomElectrodynamics}
\bibinfo{author}{Naik, R.~K.} \emph{et~al.}
\newblock \bibinfo{title}{{Random access quantum information processors using
  multimode circuit quantum electrodynamics}}.
\newblock \emph{\bibinfo{journal}{Nature Communications}}
  \textbf{\bibinfo{volume}{8}}, \bibinfo{pages}{1904} (\bibinfo{year}{2017}).

\bibitem{McKay2015High-ContrastQED}
\bibinfo{author}{McKay, D.~C.}, \bibinfo{author}{Naik, R.},
  \bibinfo{author}{Reinhold, P.}, \bibinfo{author}{Bishop, L.~S.} \&
  \bibinfo{author}{Schuster, D.~I.}
\newblock \bibinfo{title}{{High-Contrast Qubit Interactions Using Multimode
  Cavity QED}}.
\newblock \emph{\bibinfo{journal}{Physical Review Letters}}
  \textbf{\bibinfo{volume}{114}}, \bibinfo{pages}{080501}
  (\bibinfo{year}{2015}).

\bibitem{Houck2008ControllingQubit}
\bibinfo{author}{Houck, A.~A.} \emph{et~al.}
\newblock \bibinfo{title}{{Controlling the Spontaneous Emission of a
  Superconducting Transmon Qubit}}.
\newblock \emph{\bibinfo{journal}{Physical Review Letters}}
  \textbf{\bibinfo{volume}{101}}, \bibinfo{pages}{080502}
  (\bibinfo{year}{2008}).

\bibitem{Reed2010FastQubit}
\bibinfo{author}{Reed, M.~D.} \emph{et~al.}
\newblock \bibinfo{title}{{Fast reset and suppressing spontaneous emission of a
  superconducting qubit}}.
\newblock \emph{\bibinfo{journal}{Applied Physics Letters}}
  \textbf{\bibinfo{volume}{96}}, \bibinfo{pages}{203110}
  (\bibinfo{year}{2010}).

\bibitem{Gambetta2011SuperconductingCoupling}
\bibinfo{author}{Gambetta, J.~M.}, \bibinfo{author}{Houck, A.~A.} \&
  \bibinfo{author}{Blais, A.}
\newblock \bibinfo{title}{{Superconducting Qubit with Purcell Protection and
  Tunable Coupling}}.
\newblock \emph{\bibinfo{journal}{Physical Review Letters}}
  \textbf{\bibinfo{volume}{106}}, \bibinfo{pages}{030502}
  (\bibinfo{year}{2011}).

\bibitem{Rempe1987ObservationMaser}
\bibinfo{author}{Rempe, G.}, \bibinfo{author}{Walther, H.} \&
  \bibinfo{author}{Klein, N.}
\newblock \bibinfo{title}{{Observation of quantum collapse and revival in a
  one-atom maser}}.
\newblock \emph{\bibinfo{journal}{Physical Review Letters}}
  \textbf{\bibinfo{volume}{58}}, \bibinfo{pages}{353--356}
  (\bibinfo{year}{1987}).

\bibitem{Underwood2012Low-disorderPhotons}
\bibinfo{author}{Underwood, D.~L.}, \bibinfo{author}{Shanks, W.~E.},
  \bibinfo{author}{Koch, J.} \& \bibinfo{author}{Houck, A.~A.}
\newblock \bibinfo{title}{{Low-disorder microwave cavity lattices for quantum
  simulation with photons}}.
\newblock \emph{\bibinfo{journal}{Physical Review A}}
  \textbf{\bibinfo{volume}{86}} (\bibinfo{year}{2012}).

\bibitem{Allman2014TunableResonator}
\bibinfo{author}{Allman, M.} \emph{et~al.}
\newblock \bibinfo{title}{{Tunable Resonant and Nonresonant Interactions
  between a Phase Qubit and L C Resonator}}.
\newblock \emph{\bibinfo{journal}{Physical Review Letters}}
  \textbf{\bibinfo{volume}{112}}, \bibinfo{pages}{123601}
  (\bibinfo{year}{2014}).

\bibitem{Chen2014QubitCoupling}
\bibinfo{author}{Chen, Y.} \emph{et~al.}
\newblock \bibinfo{title}{{Qubit Architecture with High Coherence and Fast
  Tunable Coupling}}.
\newblock \emph{\bibinfo{journal}{Physical Review Letters}}
  \textbf{\bibinfo{volume}{113}}, \bibinfo{pages}{220502}
  (\bibinfo{year}{2014}).

\bibitem{Sirois2015Coherent-stateConversionb}
\bibinfo{author}{Sirois, A.~J.} \emph{et~al.}
\newblock \bibinfo{title}{{Coherent-state storage and retrieval between
  superconducting cavities using parametric frequency conversion}}.
\newblock \emph{\bibinfo{journal}{Applied Physics Letters}}
  \textbf{\bibinfo{volume}{106}}, \bibinfo{pages}{172603}
  (\bibinfo{year}{2015}).

\bibitem{McKay2016UniversalBusb}
\bibinfo{author}{McKay, D.~C.} \emph{et~al.}
\newblock \bibinfo{title}{{Universal Gate for Fixed-Frequency Qubits via a
  Tunable Bus}}.
\newblock \emph{\bibinfo{journal}{Physical Review Applied}}
  \textbf{\bibinfo{volume}{6}}, \bibinfo{pages}{064007} (\bibinfo{year}{2016}).

\bibitem{Lu2017UniversalQubitb}
\bibinfo{author}{Lu, Y.} \emph{et~al.}
\newblock \bibinfo{title}{{Universal Stabilization of a Parametrically Coupled
  Qubit}}.
\newblock \emph{\bibinfo{journal}{Physical Review Letters}}
  \textbf{\bibinfo{volume}{119}}, \bibinfo{pages}{150502}
  (\bibinfo{year}{2017}).

\bibitem{Halfmann1998CoherentSO2}
\bibinfo{author}{Halfmann, T.} \& \bibinfo{author}{Bergmann, K.}
\newblock \bibinfo{title}{{Coherent population transfer and dark resonances in
  SO2}}.
\newblock \emph{\bibinfo{journal}{The Journal of Chemical Physics}}
  \textbf{\bibinfo{volume}{104}}, \bibinfo{pages}{7068} (\bibinfo{year}{1998}).

\bibitem{Vasilev2009OptimumPassage}
\bibinfo{author}{Vasilev, G.~S.}, \bibinfo{author}{Kuhn, A.} \&
  \bibinfo{author}{Vitanov, N.~V.}
\newblock \bibinfo{title}{{Optimum pulse shapes for stimulated Raman adiabatic
  passage}}.
\newblock \emph{\bibinfo{journal}{Physical Review A}}
  \textbf{\bibinfo{volume}{80}}, \bibinfo{pages}{013417}
  (\bibinfo{year}{2009}).

\bibitem{Bell1964OnParadox}
\bibinfo{author}{Bell, J.}
\newblock \bibinfo{title}{{On the Einstein-Podolsky-Rosen paradox}}.
\newblock \emph{\bibinfo{journal}{Physics}} \textbf{\bibinfo{volume}{1}},
  \bibinfo{pages}{195--200} (\bibinfo{year}{1964}).

\bibitem{James2001MeasurementQubits}
\bibinfo{author}{James, D. F.~V.}, \bibinfo{author}{Kwiat, P.~G.},
  \bibinfo{author}{Munro, W.~J.} \& \bibinfo{author}{White, A.~G.}
\newblock \bibinfo{title}{{Measurement of qubits}}.
\newblock \emph{\bibinfo{journal}{Physical Review A}}
  \textbf{\bibinfo{volume}{64}}, \bibinfo{pages}{052312}
  (\bibinfo{year}{2001}).

\bibitem{Reagor2013ReachingCavities}
\bibinfo{author}{Reagor, M.} \emph{et~al.}
\newblock \bibinfo{title}{{Reaching 10 ms single photon lifetimes for
  superconducting aluminum cavities}}.
\newblock \emph{\bibinfo{journal}{Applied Physics Letters}}
  \textbf{\bibinfo{volume}{102}}, \bibinfo{pages}{192604}
  (\bibinfo{year}{2013}).

\bibitem{Wallraff2005ApproachingReadout}
\bibinfo{author}{Wallraff, A.} \emph{et~al.}
\newblock \bibinfo{title}{{Approaching Unit Visibility for Control of a
  Superconducting Qubit with Dispersive Readout}}.
\newblock \emph{\bibinfo{journal}{Physical Review Letters}}
  \textbf{\bibinfo{volume}{95}}, \bibinfo{pages}{060501}
  (\bibinfo{year}{2005}).

\bibitem{Chow2010DetectingReadout}
\bibinfo{author}{Chow, J.~M.} \emph{et~al.}
\newblock \bibinfo{title}{{Detecting highly entangled states with a joint qubit
  readout}}.
\newblock \emph{\bibinfo{journal}{Physical Review A}}
  \textbf{\bibinfo{volume}{81}}, \bibinfo{pages}{062325}
  (\bibinfo{year}{2010}).

\bibitem{Leung2017SpeedupUnits}
\bibinfo{author}{Leung, N.}, \bibinfo{author}{Abdelhafez, M.},
  \bibinfo{author}{Koch, J.} \& \bibinfo{author}{Schuster, D.}
\newblock \bibinfo{title}{{Speedup for quantum optimal control from automatic
  differentiation based on graphics processing units}}.
\newblock \emph{\bibinfo{journal}{Physical Review A}}
  \textbf{\bibinfo{volume}{95}}, \bibinfo{pages}{042318}
  (\bibinfo{year}{2017}).

\end{thebibliography}

\begin{acknowledgments}

The authors thank R. Cook, Y. Zhong, and A. A. Clerk for useful discussions, and A. Oriani for support with cryogenic facilities. This material is based upon work supported by the Army Research Office under (W911NF-15-2-0058). The views and conclusions contained in this document are those of the authors and should not be interpreted as representing the official policies, either expressed or implied, of the Army Research Laboratory or the U.S. Government. The U.S. Government is authorized to reproduce and distribute reprints for Government purposes notwithstanding any copyright notation herein. Research was also supported by the U. S. Department of Defense under DOD contract H98230-15-C0453. This work was partially supported by the University of Chicago Materials Research Science and Engineering Center, which is funded by the National Science Foundation
under Award No. DMR-1420709. This work made use of the Pritzker
Nanofabrication Facility of the Institute for Molecular
Engineering at the University of Chicago, which receives support from SHyNE, a node of the National Science Foundation's National Nanotechnology Coordinated Infrastructure (NSF NNCI-1542205). We gratefully acknowledge support from the David and Lucile Packard Foundation.
\end{acknowledgments}

\section*{Author contributions}
N.L., Y.L designed and fabricated the device, designed the experimental protocols, performed the experiments, and analyzed the data. K.J. provided theoretical support. S.C., R.N., N.E., R.M provided fabrication and experimental support. All authors co-wrote the paper.

\section*{Competing financial interests}
The authors declare no competing financial interests.

\clearpage
\newpage

\appendix
\onecolumngrid
\section{Cryogenic setup and control instrumentation}

\begin{figure}[h!]
  \begin{center}
    \includegraphics[width= \textwidth]{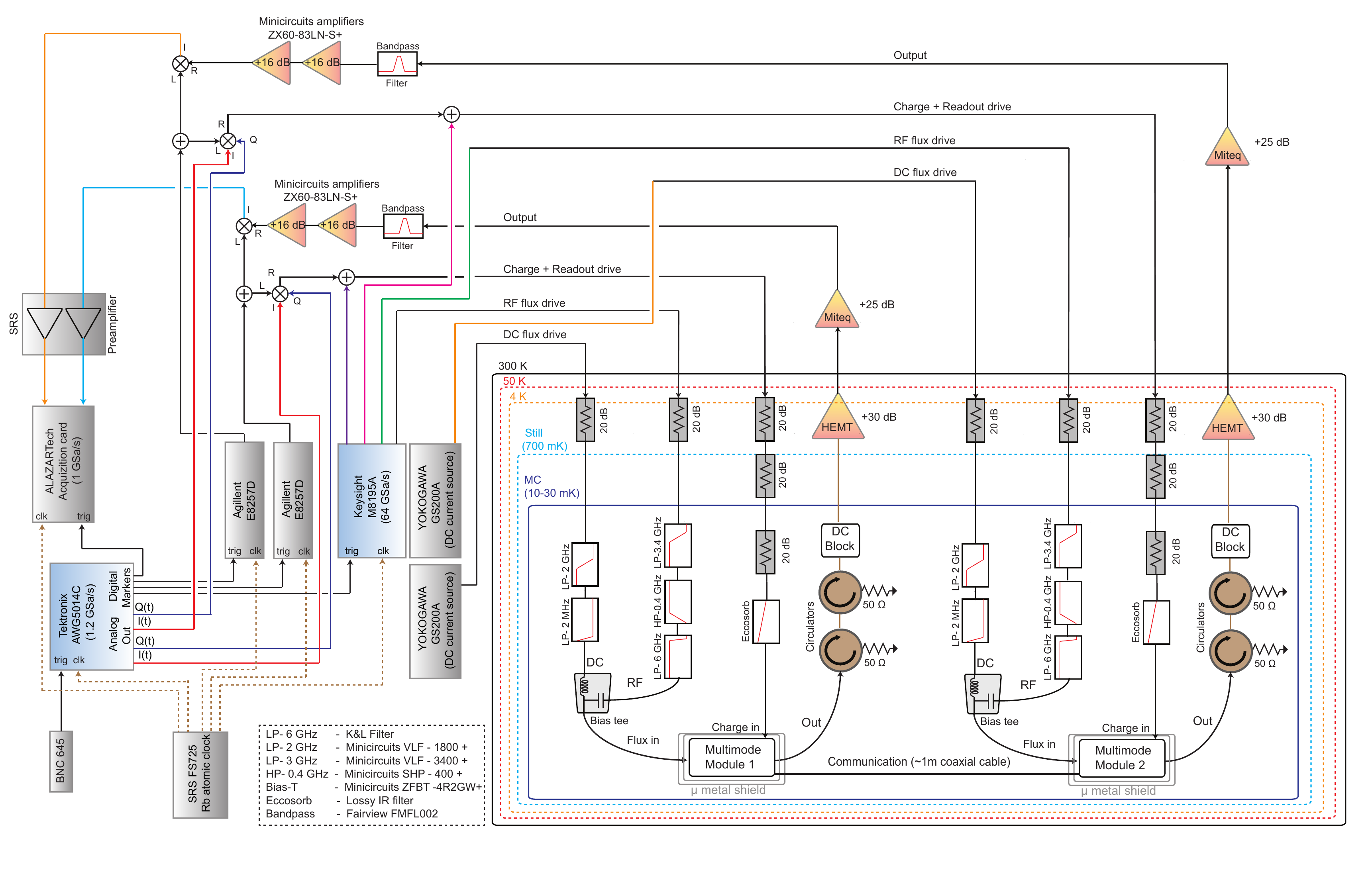}

    \caption{Detailed schematic of the cryogenic setup, control instrumentation, and the wiring of microwave and DC connections to the device.}
  	\label{wiring diagram}
  \end{center}
\end{figure} 

The device is heat sunk via an OFHC copper post to the base stage of a Bluefors dilution refrigerator (10-30 mK). The sample is surrounded by a can containing two layers of $\mu$-metal shielding and a layer of lead shielding, thermally anchored using an inner close fit copper shim sheet, attached to the copper can lid. 
The schematic of the cryogenic setup, control instrumentation, and the wiring of the device is shown if Supplementary Figure \ref{wiring diagram}. Each device is connected to the rest of the setup through three ports: a charge port that applies qubit and readout drive tones, a flux port for shifting the qubit frequency using a DC-flux bias current and for applying RF sideband flux pulses, and an output port for measuring the transmission from the readout resonator. The readout pulses are generated by mixing a local oscillator tone (generated from an Agilent 8257D RF signal generator), with pulses generated by a Tektronix AWG5014C arbitrary waveform generator (TEK) with a sampling rate of 1.2 GSa/s, using an IQ-Mixer (MARQI MLIQ0218). The charge drive pulses are generated with Keysight M8195A arbitrary waveform generator by direct synthesis, and subsequently combined with the readout drive pulse. The combined signals are sent to the device, after being attenuated a total of 60 dB in the dilution fridge, using attenuators thermalized to the 4K (20 dB), still (20 dB) and base stages (20 dB). The charge drive line also includes a lossy ECCOSORB CR-117 filter to block IR radiation, and a low-pass filter with a sharp roll-off at 6\,GHz, both thermalized to the base stage. The flux-modulation pulses are also directly synthesized by the Keysight M8195A arbitrary waveform generator and attenuated by $20$ dB at the 4 K stage, and bandpass filtered to within a band of 400\,MHz - 3.4\,GHz at the base stage, using the filters indicated in the schematic. The DC flux bias current is generated by a YOKOGAWA GS200 low-noise current source, attenuated by 20 dB at the 4 K stage, and low-pass filtered down to a bandwidth of 1.9\,MHz. The DC flux bias current is combined with the flux-modulation pulses at a bias tee thermalized at the base stage. The state of the transmon is measured using the transmission of the readout resonator, through the dispersive circuit QED readout scheme \cite{Wallraff2005ApproachingReadout}. The transmitted signal from the readout resonator is passed through a set of cryogenic circulators (thermalized at the base stage) and amplified using a HEMT amplifier (thermalized at the 4 K stage). Once out of the fridge, the signal is filtered (narrow bandpass filter around the readout frequency) and further amplified. The amplitude and phase of the resonator transmission signal are obtained through a heterodyne measurement, with the transmitted signal demodulated using an IQ mixer and a local oscillator at the readout resonator frequency. The heterodyne signal is amplified (SRS preamplifier) and recorded using a fast ADC card (ALAZARtech).

\section{Device Hamiltonian}

Without connecting to the coaxial cable, the Hamiltonian of the i-th (i=1,2) circuit can be modeled by

\begin{align}
\hat{H} = &h\nu_{i,q} (t) \hat{a}_{i}^\dagger \hat{a}_{i} +\frac{1}{2}\alpha_{i} \hat{a}_{i}^{\dagger} \hat{a}_{i} (\hat{a}_{i}^{\dagger} \hat{a}_{i} -1) + h\nu_{i,r} \hat{b}_{i,r}^\dagger \hat{b}_{i,r} + h\nu_{i,c} \hat{b}_{i,c}^\dagger \hat{b}_{i,c} + \sum_{m=1}^8 h\nu_{i,m} \hat{b}_{i,m}^\dagger \hat{b}_{i,m} \nonumber \\ 
&+ h g_{i,r}(\hat{b}_{i,r} + \hat{b}^{\dagger}_{i,r})(\hat{a}_{i} + \hat{a}_{i}^\dagger)  + h g_{i,c}(\hat{b}_{i,c} + \hat{b}^{\dagger}_{i,c})(\hat{a}_{i} + \hat{a}_{i}^\dagger)  + \sum_{m=1}^{8} h g_{i,m}(\hat{b}_{i,m} + \hat{b}^{\dagger}_{i,m})(\hat{a}_{i} + \hat{a}_{i}^\dagger) \label{eq:H_chip}
\end{align}

where $\hat{a}_{i}$, $\hat{b}_{i,r}$, $\hat{b}_{i,c}$ and $\hat{b}_{i,m}$ stand for the annihilation operators of the flux-tunable qubit, the readout resonator, the communication cavity and the m-th multimode on the i-th chip. The communication cavities of the two chips are of identical coplanar waveguide resonator design with large center pin and gap width, leading to approximately the same resonant frequency $\nu_{1,c} \approx \nu_{2,c} = \nu_c$ and the same coupling strength $g_l$ to the coaxial cable mode $\hat{b}_l$,

\begin{align}
\hat{H}_{int} = \sum_{i=1}^{2} h\nu_{c} \hat{b}_{i,c}^\dagger \hat{b}_{i,c} + h\nu_{l} \hat{b}_{l}^\dagger \hat{b}_{l} +\sum_{i=1}^{2} h g_{l}(\hat{b}_{l} \hat{b}^{\dagger}_{i,c}+ \hat{b}^{\dagger}_{l} \hat{b}_{i,c}).
\label{eq:H_int}
\end{align}

\ref{eq:H_int} can be directly diagonalized, yielding three normal modes $\tilde{\hat{b}}_{1}$, $\tilde{\hat{b}}_{2}$ and $\tilde{\hat{b}}_{c}$,

\begin{align}
\tilde{\hat{H}}_{int} = h\nu_{c} \tilde{\hat{b}}_{c}^\dagger \tilde{\hat{b}}_{c}+h\nu_{1} \tilde{\hat{b}}_{1}^\dagger \tilde{\hat{b}}_{1}+h\nu_{2} \tilde{\hat{b}}_{2}^\dagger \tilde{\hat{b}}_{2},
\label{eq:H_bar_int}
\end{align}
where
\begin{align}
\nu_{1}&=\nu_{c}+\frac{\delta}{2}+\sqrt{8g_l^2+\delta^2},\nonumber\\
\nu_{2}&=\nu_{c}+\frac{\delta}{2}-\sqrt{8g_l^2+\delta^2},\label{eq:nu}
\end{align}
and

\begin{align}
\tilde{\hat{b}}_{c}&=\frac{1}{\sqrt{2}}(\hat{b}_{1,c}-\hat{b}_{2,c}),\nonumber\\
\tilde{\hat{b}}_{1}&=\frac{1}{\sqrt{2+(r+\sqrt{2+r^2})^2}}(\hat{b}_{1,c}+\hat{b}_{2,c}+(r+\sqrt{2+r^2})\hat{b}_l),\nonumber\\
\tilde{\hat{b}}_{2}&=\frac{1}{\sqrt{2+(r-\sqrt{2+r^2})^2}}(\hat{b}_{1,c}+\hat{b}_{2,c}+(r-\sqrt{2+r^2})\hat{b}_l).
\label{eq:bhats}
\end{align}
Here $\delta$ stands for the deviation of the cable mode frequency from the communication resonator frequency, i.e. $\delta = \nu_l-\nu_c$, and $r=\delta/2g_l$. The normal mode frequencies relative to the qubit frequency can be readily obtained from Fig.~\ref{comm_coherence}.b, so that $\delta$ and $g_l$ can be calculated from Eq.~\ref{eq:H_bar_int} and \ref{eq:nu}. Eq.~\ref{eq:H_chip} and \ref{eq:bhats} together give the renormalized coupling strengths between the qubit and these normal modes,

\begin{align}
\tilde{\hat{g}}_{c}&=\frac{g_c}{\sqrt{2}},\nonumber \\
\tilde{\hat{g}}_{1}&=\frac{g_c}{\sqrt{2+(h+\sqrt{2+h^2})^2}},\nonumber\\     
\tilde{\hat{g}}_{2}&=\frac{g_c}{\sqrt{2+(h-\sqrt{2+h^2})^2}}.
\label{eq:g_normal}
\end{align}

It is worth noting that the center normal mode, $\tilde{\hat{b}}_{c}$, is selected to be our communication channel mode in the experiment, for two obvious reasons: it contains only the two resonator modes with no convolution with the cable mode, as seen in Eq.~\ref{eq:bhats}, thus highly immune to the photon loss of the cable. Eq.~\ref{eq:g_normal} shows that it also couples more strongly to the qubit comparing to the other two normal modes, which also agrees well with Fig.~\ref{comm_coherence}.b where the center chevron has the fastest oscillation. 

\begin{figure}[H]
  \centering \includegraphics[width=0.7\columnwidth]{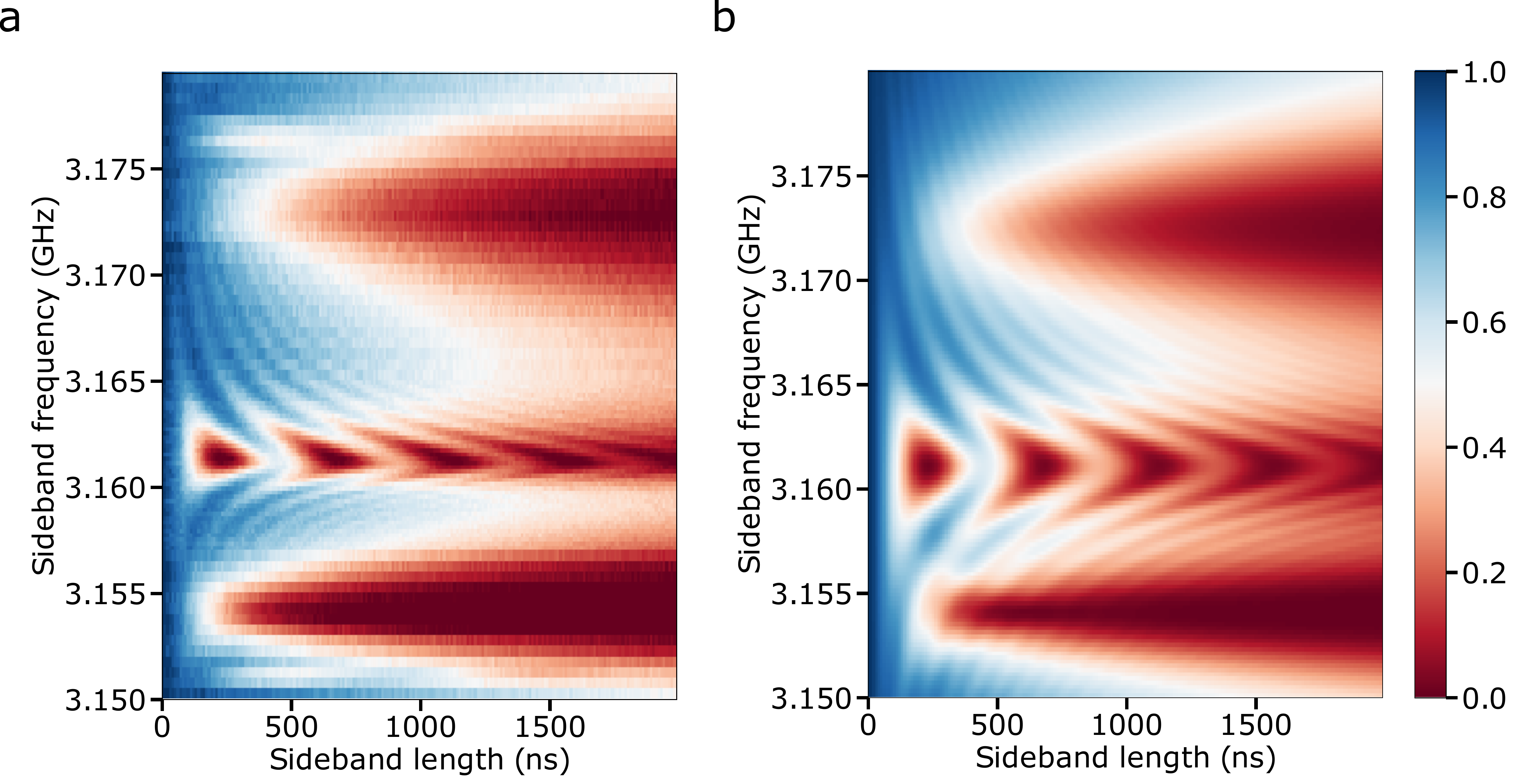}
  \caption{{\textbf{Stimulated vacuum Rabi oscillation between the qubit and the communication mode.} By fitting to the experimental data (a) using our analytical model, we extracted the deviation of the cable mode frequency from the two communication resonators to be 4.25~MHz, while the coupling between the cable mode and the communication resonator is $6.46$~MHz. Plugging these along with other circuit parameters obtained from the experiment into a master equation, we can simulate the experimental result with decent agreement (b). }}
\label{comm_coherence_sim}
\end{figure}

Fitting Eq.~\ref{eq:nu} to Fig.~\ref{comm_coherence}b, we obtain $\delta= 4.25$MHz and $g_l= 6.46$MHz, from which we can numerically reproduce the chevron patterns observed in the experiment (Fig.\ref{comm_coherence_sim}).

Here we list the relevant circuit parameters in the following table:

\begin{center}
\renewcommand{\arraystretch}{1.5}
\begin{tabular}{ |c|c|c| } 
\hline
  & sample 1 & sample 2  \\ 
 \hline
 $\nu_q$ & static: 4.7685\,GHz; range: $\approx$ 3.0 - 5.9\,GHz & static: 4.7420\,GHz; range: $\approx$ 3.5 - 5.5\,GHz  \\ 
 \hline
 $\alpha$ & 109.8\,MHz& 109.9\,MHz\\ 
 \hline
 $\nu_r$ & 5.7463\,GHz& 5.7405\,GHz  \\ 
 \hline
 $\nu_c$ & $\approx$ 7.88\,GHz& $\approx$ 7.88\,GHz  \\ 
 \hline
 $\nu_m$ & 5.9 - 7.6\,GHz& 5.9 - 7.6\,GHz  \\ 
 \hline
  $T_1$ & 10.1\,$\mu$ s& 7.9\,$\mu$ s  \\ 
 \hline
  $T_2^*$ & 0.7\,$\mu$ s& 1.4\,$\mu$ s  \\ 
 \hline
\end{tabular}
\end{center}


\section{Sideband interaction and calibrations}
\begin{figure}[H]
  \centering \includegraphics[width=0.7\columnwidth]{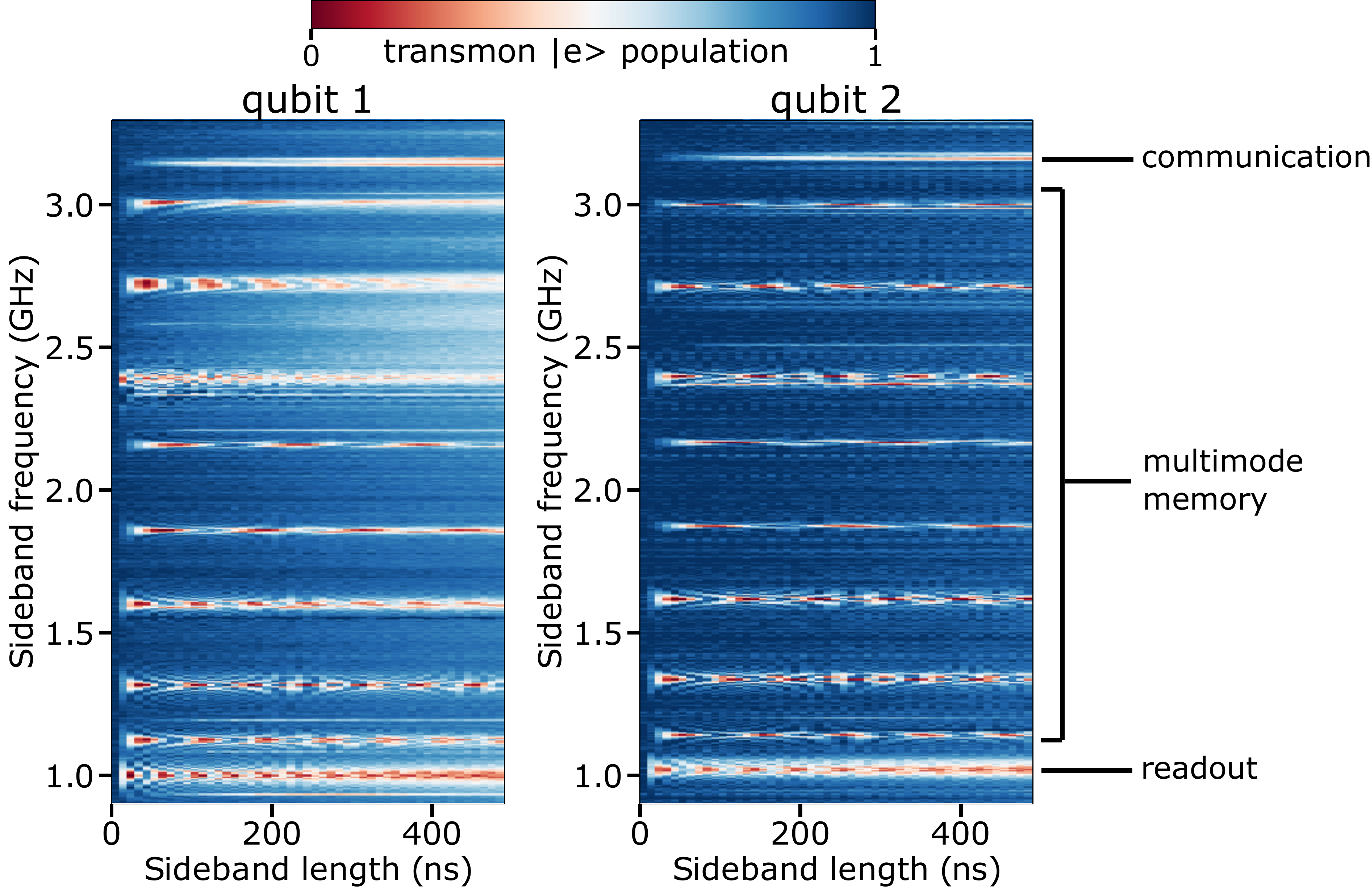}
  \caption{{\textbf{Full sideband Rabi spectrum of each qubit. } Stimulated vacuum Rabi oscillation with sideband frequency scan covering the band of all resonance frequencies of the resonators. The clean chevron patterns indicate that our transmons are free from spurious crosstalks. We can clearly identify ten chevron patterns corresponding to one readout resonator (lowest frequency), eight multimode memory resonators and one communication resonator (highest frequency).}}
\end{figure}

In these scans, we can clearly identify ten chevron patterns corresponding to one readout resonator, eight multimode memory resonators, and one communication resonator. The crosstalk at sideband frequency $\approx$ 2.4\,GHz corresponds to directly driving the g-e transition at the half of this frequency. The clean chevron patterns indicate that our transmons are free from spurious crosstalks. Compared to the segmented scans in the main text, these sideband scans are taken at higher amplitude to broaden the chevron patterns for better visualization. The chevron patterns also show with faster oscillations and slightly higher frequency due to DC-offset, described in the next section.


The essential ingredient of photonic communication for our devices is the flux sideband interaction. It is therefore important to calibrate the sideband interactions well on both devices for obtaining high fidelity photonic communication. For our devices, this involves using the correct amplitude, frequency and timing of the sideband interactions. This section describes our calibration protocols for these parameters.

\begin{figure}[H]
  \centering \includegraphics[width=0.7\columnwidth]{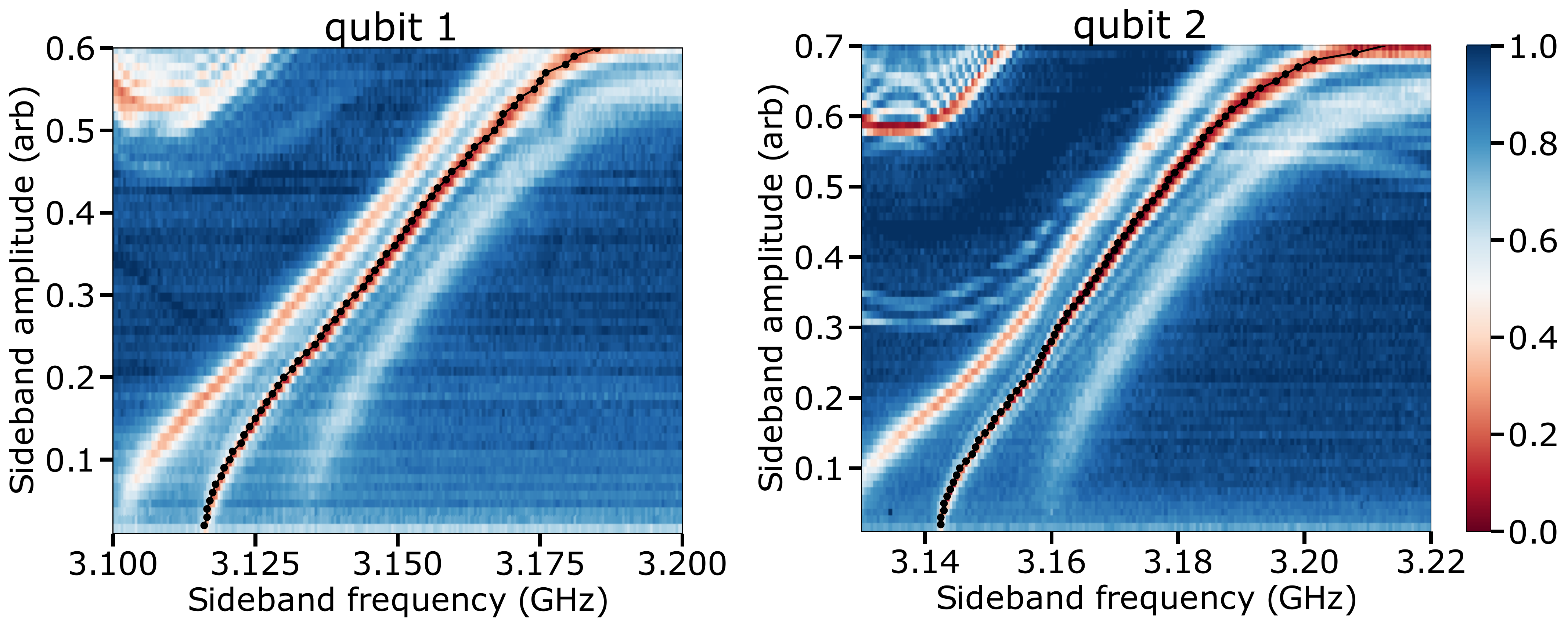}
  \caption{{\textbf{DC offset scan. }There is a shift (DC-offset) of the qubit frequency during the flux modulation, arising from the non-linear flux-frequency relation of the transmon. To calibrate this effect, we sweep sideband transition frequency at different flux amplitudes and obtain the calibration with linear interpolation. The black dots on the figures show the tracked resonance sideband frequency for the considered range of amplitude. The pattern of three normal modes persisted for the considered range of sideband amplitudes. In this experiment, we set the sideband length to be inversely proportional to the sideband amplitude. This ensures high contrast features even for small sideband amplitude which the coupling is weak.}}
\label{dc_offset}
\end{figure}

First, we run a 2D sweep (sideband amplitude and sideband frequency) of stimulated vacuum rabi around the communication mode frequencies. The main feature in figure \ref{dc_offset} shows a clear pattern of three resonances, corresponds to the three hybridized normal modes of the communication channel. From the data, it is obvious that the resonance frequencies are dependent on the sideband amplitudes. The effect originated from the non-linear flux-frequency relation of the transmon, causing a shift (DC-offset) of the qubit frequency during the flux modulation. By doing a finer scan around the resonance frequency, we calibrated the on-resonance frequency of each sideband amplitude with an accuracy of 100 kHz.

\begin{figure}[H]
  \centering \includegraphics[width=0.7\columnwidth]{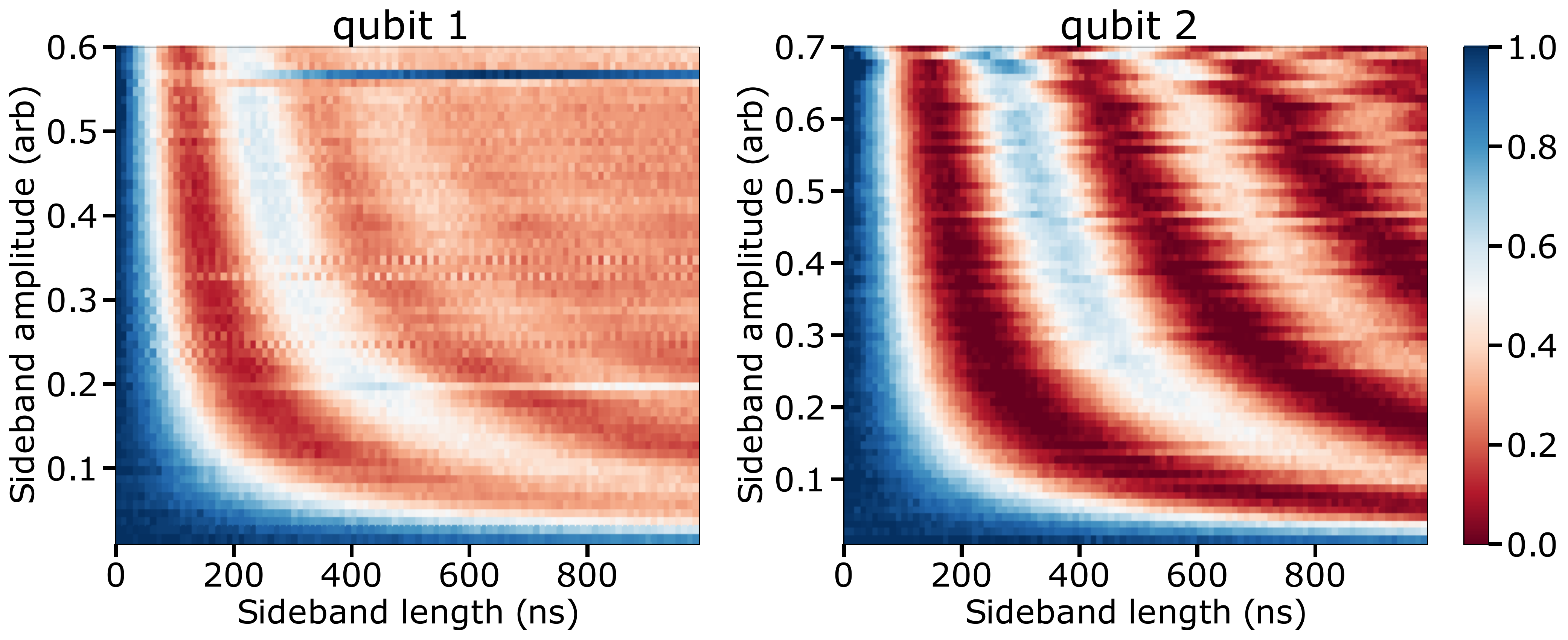}
  \caption{{\textbf{Sideband Rabi sweep. }Using the calibrated DC offset, we obtained sideband rabi data between transmon and communication resonator for different sideband amplitude. Notice that the contrast of qubit 1 is much smaller than qubit 2. This is because qubit 2 has a higher coherence time. The trajectories of these scans are used for fitting the effective qubit decay parameters during the sideband coupling. These decay parameters are subsequently being applied in master equation simulation of photon transfer and Bell entanglement generation. }}
\label{comm_sideband_rabi}
\end{figure}

With the calibrated frequencies, we sweep the sideband length with a range of sideband amplitudes and obtain stimulated vacuum Rabi oscillation. The experimental data is displayed in figure \ref{comm_sideband_rabi}. As expected, a higher sideband amplitude implies a higher effective coupling rate. Using this data, we obtained the effective qubit dissipation parameters during the sideband coupling. These dissipation parameters are subsequently being applied in master equation simulation of photon transfer and Bell entanglement generation.

\begin{figure}[H]
  \centering \includegraphics[width=0.4\columnwidth]{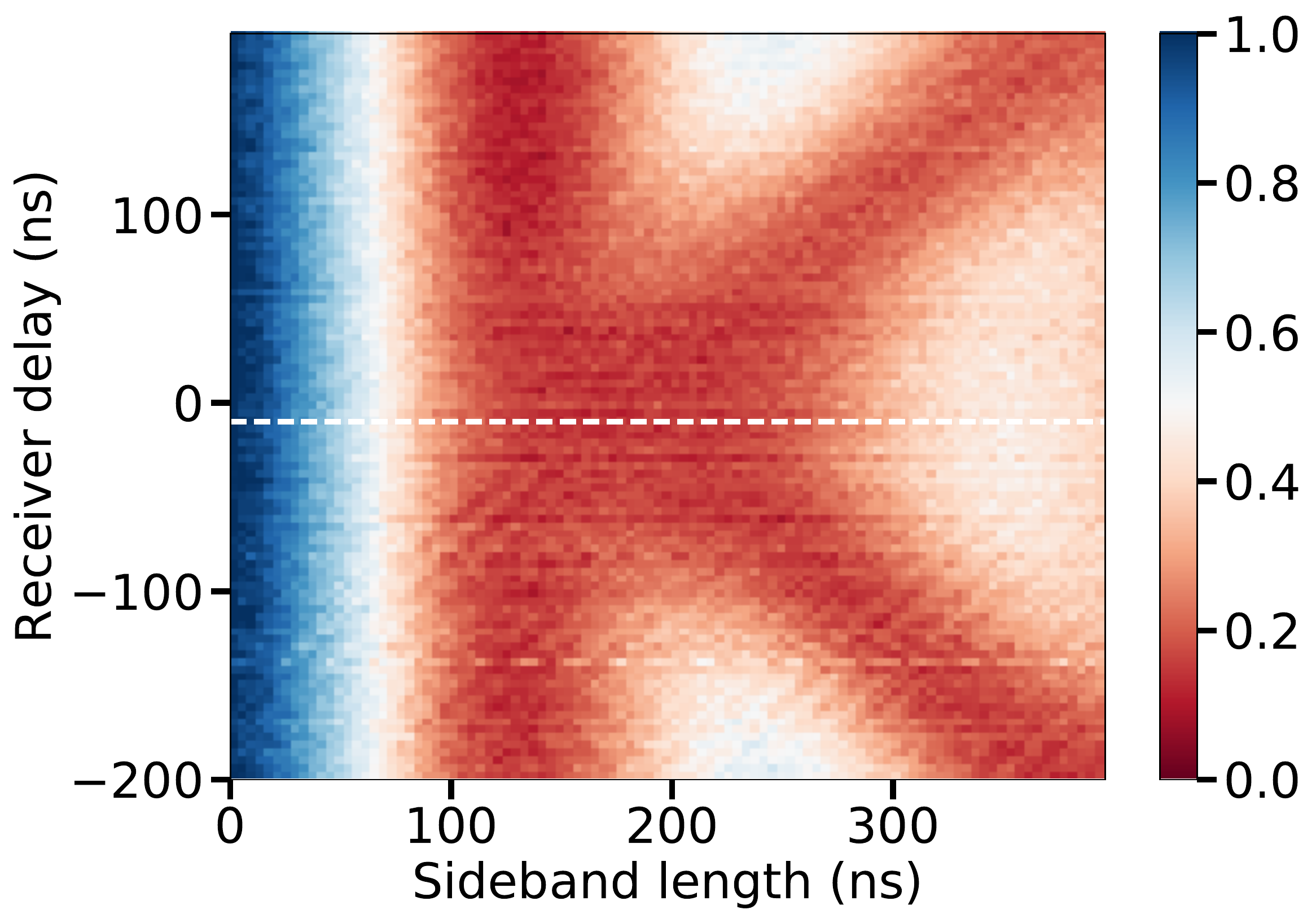}
  \caption{{\textbf{Delay calibration. } This figure shows the population of the sender qubit with sweeping parameters of two sideband length and receiver delay. Ideally, the two flux sidebands during photon transfer should start simultaneously. However due to experimental conditions (e.g. different travel path length of flux line control from AWG to sample) causes the sideband pulses start at a different time on the devices, even the AWG is programmed to initiate two pulses simultaneously. To calibrate this effect, we sweep the delay between two sideband pulses and found that the flux control of qubit 2 is delayed by 10\,ns. Throughout the experiment we time-advanced the control of qubit 2 flux by 10\,ns in our pulse generation software.}}
\label{delay_cal}
\end{figure}

Lastly, we calibrated the timing of the two flux sideband pulses. Due to slightly different travel path length of flux line control from AWG to sample, we expect a slightly different timing between the two flux sideband pulses. Since the simultaneity of two flux sideband pulses is essential for high fidelity transfer, it is important to calibrate this systematic error. The experiment was conducted with two equal length sideband pulses but sweeping the software delay between two pulses. Here, a negative receiver delay means the sender qubit (qubit 1) sideband pulse starts before the receiver qubit (qubit 2) sideband pulse. Figure \ref{delay_cal} shows the population of the sender qubit with sweeping parameters of two sideband length and receiver delay. The center of the ``K" pattern corresponds to the scenario where the photon is maximally captured by the receiver qubit. We obtained the ``K" pattern as symmetric around receiver delay time of $\approx$ -10\,ns, indicating the flux sideband pulse of the receiver qubit (qubit 2) lags the flux sideband pulse of the sender qubit (qubit 1). As a sanity check, we switched the role of sender and receiver qubit, such that sender is qubit 2 and receiver is qubit 1. In such case, we found that the pattern is symmetric around receiver delay time of $\approx$ +10\,ns. This confirms our conclusion that indeed the qubit 2 lags the flux sideband pulse of qubit 1 due to a delay in the lines.

\section{Master equation simulation}
In order to calculate the communication processes between the remote qubits using master equation simulations, we first write out the circuit Hamiltonian under flux modulations, based on Eq.~\ref{eq:H_chip}$\sim$\ref{eq:g_normal}, as

\begin{align}
\hat{H} = &\sum_{i=1}^2 \sum_{j=1}^3  h(\nu_{i,q} + \epsilon_i \cos \omega_i t) \hat{a}_{i}^\dagger \hat{a}_{i} +h\alpha_{i} \hat{a}_{i}^{\dagger 2} \hat{a}_{i}^2 + h\nu_{j} \hat{b}_{j}^\dagger \hat{b}_{j} + h g_{j}(\hat{b}_{j} + \hat{b}^{\dagger}_{j})(\hat{a}_{i} + \hat{a}_{i}^\dagger), \label{eq:H_t}
\end{align}
where $\hat{b}_{j}$ stand for the three normal mode and $g_j$ their coupling strengths to the two transmon qubits. Assuming weak flux modulation with $\omega_i \approx \nu_c-\nu_{i,q}$, and under the rotating frame transformation $U=$exp$[-ih \sum_{i=1}^2 \sum_{j=1}^3 ( (\nu_{i,q}t - \frac{\epsilon_i}{2\omega_i} \cos \omega_i t) \hat{a}_{i}^\dagger \hat{a}_{i} + \nu_{c} \hat{b}_{j}^\dagger \hat{b}_{j} t)]$, Eq.~\ref{eq:H_t} can be rewritten as

\begin{align}
\hat{H} = \sum_{i=1}^2 \sum_{j=1}^3 \left\{h\alpha_{i} \hat{a}_{i}^{\dagger 2} \hat{a}_{i}^2+ h(\nu_j-\nu_c) \hat{b}_{j}^\dagger \hat{b}_{j}- ih g_{j} J_{1}\left(\frac{\epsilon_i}{2\omega_i}\right)\left[\hat{b}_{j} \hat{a}_{i}^\dagger e^{i(\omega_i-\nu_j-\nu_c)t}-\hat{b}_{j}^\dagger \hat{a}_{i} e^{-i(\omega_i-\nu_j-\nu_c)t}\right] \right\}. \label{eq:H_rot}
\end{align}

Here $J_1(x)$ stands for the Bessel function of the first kind of the first order, and all the fast-oscillating terms have been abandoned. With the flux-modulation frequencies being $\omega_i = \nu_c-\nu_{i,q}$, and applying the two-level-approximation for the qubits, we find the "transfer Hamiltonian" as

\begin{align}
\hat{H} =  \sum_{i=1}^2 \sum_{j=1}^2 h(\nu_{l,j}-\nu_c) \hat{b}_{l,j}^\dagger \hat{b}_{l,j} - ih J_{1}\left(\frac{\epsilon_i}{2\omega_i}\right) \left[g_{l,j} \left(\hat{b}_{l,j} \hat{\sigma}_{i}^+-\hat{b}_{l,j}^\dagger \hat{\sigma}_{i}^-\right) + g_{c} \left(\hat{b}_{c} \hat{\sigma}_{i}^+-\hat{b}_{c}^\dagger \hat{\sigma}_{i}^-\right)\right], \label{eq:H_trans}
\end{align}
where $\hat{b}_{l,1}$ and $\hat{b}_{l,2}$ are the two lossy ``bright'' normal mode, and $b_c$ is the ``dark" communication channel mode. Plugging this into the master equation,

\begin{equation}
\dot{\rho}=-i[\hat{H},\rho]+\kappa_{l,j} \mathcal{D}[\hat{b}_{l,j}]\rho +\kappa_{c} \mathcal{D}[\hat{b}_{c}]\rho + \gamma \mathcal{D}[\hat{\sigma}_-]\rho + \gamma_{\phi} \mathcal{D}[\hat{\sigma}_z]\rho,\label{eq:master equation}
\end{equation}
we are able to simulate the bidirectional photon transfer experiment (Fig.~\ref{two_way}) and the remote entanglement experiment (Fig.~\ref{Bell}). 

\begin{figure}[H]
  \centering \includegraphics[width=0.8\columnwidth]{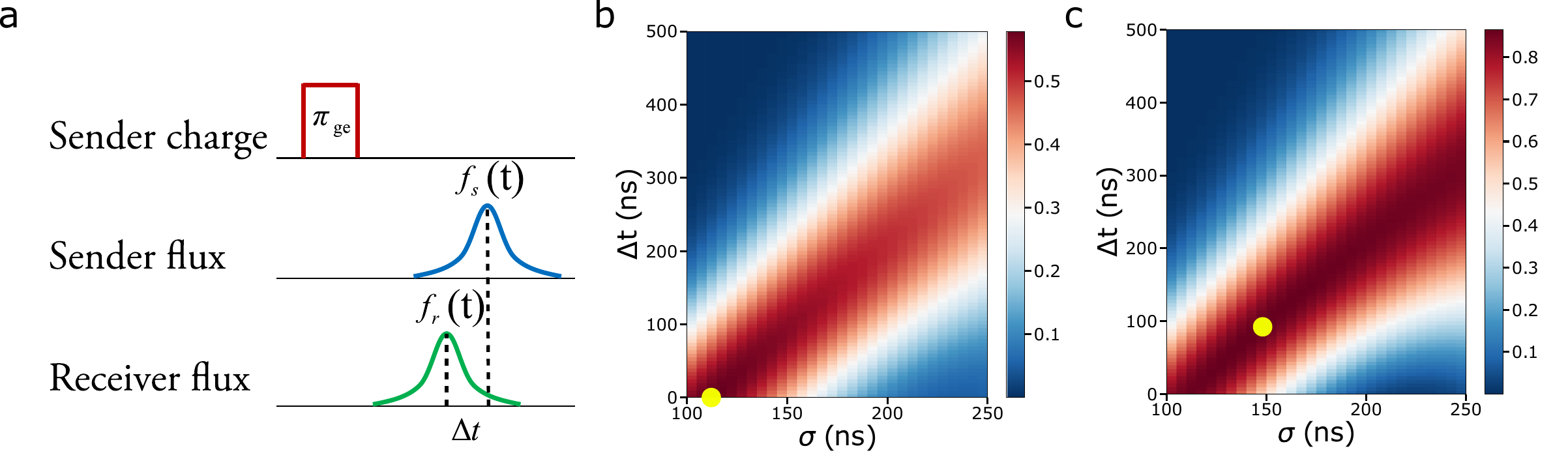}
  \caption{{\textbf{STIRAP-like protocol.} (a) Pulse sequence of the STIRAP protocol for photon transfer. After initializing of the sender qubit state in the excited state, two Gaussian pulses with same duration and amplitude (set to be the maximum amplitude achievable in the experiment) are applied to the flux channels of the two qubits, with the receiver pulse turned on ahead of the sending pulse by a time of $\Delta t$. (b) Calculation of the transfer fidelity as a function of the Gaussian RMS width, $\sigma$, as well as the delay time $\Delta t$. A maximum fidelity of $56\%$ occurs at $\{\sigma = 120~us,\Delta t = 0~us\}$ (labeled by the yellow dot), which is worse than the 60\% fidelity achieved by the simultaneous square pulse scheme. This indicates that, in our current parameter regime, the fidelity is optimal with simultaneous square pulse scheme which has the shortest pulse length. (c) With better qubit coherence properties of $T1,T2=20$~us, the STIRAP protocol promises 85\% maximum fidelity at $\{\sigma = 145~us,\Delta t = 95~us\}$ (labeled by the yellow dot), which is higher than the maximum fidelity of 82\% yielded by the simultaneous square pulse scheme under the same parameters. 
}}
\label{stirap}
\end{figure}

Simultaneous square sideband pulses are adopted in both the photon transfer and Bel state creation experiment to achieve the shortest pulse time possible, as is shown in fig.~\ref{two_way} and \ref{Bell}. However, there is a possibility that better fidelities could be acquired through further minimizing the photon loss in the communication mode, by making use of adiabatic protocols in a manner akin to the stimulated Raman adiabatic passage (STIRAP). A typical STIRAP protocol has a pulse sequence shown in fig.~\ref{stirap}a, where after the excitation of the sender qubit, the receiving pulse turns on first, and slowly ramps down together with the ramping up of the sending pulse. When the ramping of the pulses are done adiabatically w.r.t the gap between the communication mode and the qubit modes, the transfer could be completed without inducing the communication mode population, and is therefore immune to the photon loss in the communication mode. However, this comes at the cost of much longer transfer time, which introduces more loss from the qubits. 

For simplicity we model the sender and receiver pulses as two Gaussian pulses with the same maximum amplitude as the square pulse scheme used in our experiment. In the time domain, the two pulses are set to be

\begin{equation}
f_s (t) = 
\left\{\begin{matrix}
A e^{-\frac{(t-t_0)^2}{2\sigma^2}}, &\left|t-t_0\right| \leqslant 5\sigma \\
0, &\left|t-t_0\right| > 5\sigma
\end{matrix}\right. ,\quad
f_r (t) = 
\left\{\begin{matrix}
A e^{-\frac{(t-t_0-\Delta t)^2}{2\sigma^2}}, &\left|t-t_0-\Delta t\right| \leqslant 5\sigma \\
0, &\left|t-t_0-\Delta t\right| > 5\sigma
\end{matrix}\right..
\label{eq:stirap_pulse}
\end{equation}
The fidelity yielded by this protocol is calculated as a function of both the pulse width $\sigma$ and the delay time $\Delta t$, via master equation simulation with real circuit parameters. Fig.~\ref{stirap}b shows that a maximum fidelity of 56\% is achieved when two Gaussian pulses with $\sigma = 120$~us overlap each other, which indicates that non-adiabatic transfer with shortest time is favorable in our current parameter regime. This also justifies our choice of the simultaneous square pulse scheme which is the fastest in all non-adiabatic schemes. In contrast, if the coherence of the qubit is improved to $T_1 = 20$~us and $T_2 = 20$~us, the same simulation results in a maximum fidelity of 85\% at delay time $\Delta t=$ (fig.~\ref{stirap}c) that is higher than the simultaneous square pulse fidelity of 82\%, proving the usefulness of the adiabatic protocol for future improvements.

\section{Readout and state tomography}


To measure the two-qubit state, we record the homodyne voltage for each qubit from every run. For example, run $i$ of the experiment would result in a 4D heterodyne voltage
values ($V_{I1,i}$, $V_{Q1,i}$, $V_{I2,i}$, $V_{Q2,i}$). These voltages are random numbers generated from a specific distribution corresponding to state projection and experimental noise. To measure
the population in the four two-qubit basis states: $|gg\rangle,|ge\rangle,|eg\rangle,|ee\rangle$ we construct the histograms for these
states by applying $\pi$ pulses to the qubits. These histograms
approximate the probability distribution for measuring a given voltage pair when the system is in a given basis state. 

We employed logistic regression for classification of the two-qubit states. By setting decision thresholds for maximizing the classification accuracy for the two-qubit basis states according to the voltage distribution, we obtain a confusion matrix representing the correct and incorrect identification of basis state. For an unknown density matrix $\rho$ we construct the classification distribution for $\rho$ from N measurements, and project onto the basis states by applying the inverse of the calculated confusion matrix. 

We perform state tomography using the standard method by calculating the linear estimator,

\begin{equation} 
\label{H_cable_normal}
\begin{split}
\rho_{est} = \sum_{i,j} \frac{Tr[(\sigma_i \otimes \sigma_j)\rho](\sigma_i \otimes \sigma_j)}{4}
\end{split}
\end{equation}

To calculate the term $Tr[(\sigma_i \otimes \sigma_j)\rho$ we apply a unitary operator $U$ to $\rho$ prior to measurement. For two-qubits, there are nine required measurements corresponding to the following unitary operators, $(I, R_Y(\pi/2), R_X(\pi/2)) \otimes (I, R_Y(\pi/2), R_X(\pi/2))$.

This simple linear estimator method can return unphysical results because it projects onto the space of all Hermitian matrices with Trace 1. However a physical density matrix must also be positive semi-definite. Following the maximum likelihood protocol outlined in \cite{McKay2015High-ContrastQED,James2001MeasurementQubits}, we estimate the most likely physical density matrix by minimizing the function,

\begin{equation} 
\begin{split}
F[\rho_{est}] = \sum_{i=1,j=1}^{N,4} (\langle j|U^\dagger_i \rho_{est} U_i |j \rangle - P_{i,j})^2
\end{split}
\end{equation}, where $U_i$ are the set of $N$ applied tomography pulses, $|j\rangle$ is the $j^{th}$ basis state, $P_{i,j}$ is the measured probability, and $\rho_{est}$ is a physical density matrix satisfying the physical constraints. The starting guess for the minimization is the density matrix estimated from the linear estimator with all negative eigenvalues set to zero. To form a over-complete set for a total of 17 tomography measurements, we also measure the negative pulse set \cite{Chow2010DetectingReadout} $(I, R_Y(-\pi/2), R_X(-\pi/2)) \otimes (I, R_Y(-\pi/2), R_X(-\pi/2))$.

\section{Online Gaussian process for Bell state optimization}

\begin{figure}[H]
  \centering \includegraphics[width=0.5\columnwidth]{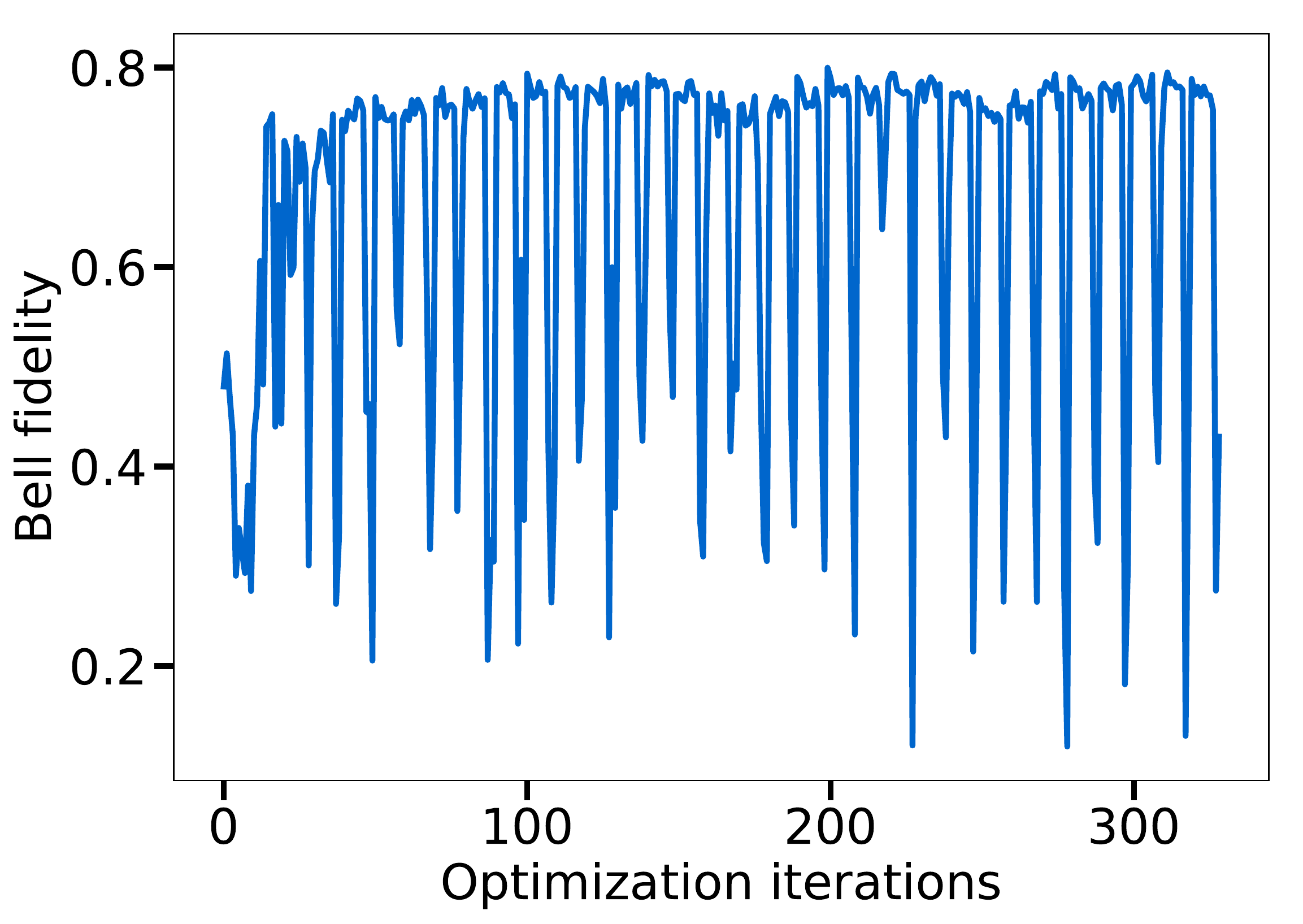}
  \caption{{\textbf{Optimization of Bell state creation with an online Gaussian process.} We employed an online optimization directly applying on the experimental device. In each iteration, the Gaussian process model proposes 8 candidate solution (1 obtained from L-BFGS-B optimization on the Gaussian model, and 7 obtained from random sampling filtered with the best model prediction), and we also test two candidate solutions from pure random sampling to improve parameter space exploration. The random samplings lead to the apparent spikes of low fidelity Bell state during the optimization iterations. The model quickly starts to converge, and after some time we obtained Bell state with a fidelity close to 80\%.}}
\label{bell_optimization}
\end{figure}

For two square pulses, there are in total 6 parameters (amplitude, frequency, and duration of each square pulse). The linear interpolation calibration of the DC offsets relates the amplitude and frequency parameters, thus resulting in 4 parameters to be optimized. All the parameters are fairly dependent on each other in the process of simultaneous transfer, meaning all 4 parameters have to be optimized together. Exhaustive search is quite forbidden even with just 4 parameters. Therefore, we employed optimization techniques, in particular, the Gaussian process to assist in optimizing Bell state creation. We employed an online optimization directly applied to the experimental device. In each iteration, the Gaussian process model proposes 8 candidate solution (1 obtained from L-BFGS-B optimization on the Gaussian model, and 7 obtained from random sampling filtered with the best model prediction), and we also test 2 candidate solution from pure random sampling to improve parameter space exploration. Figure \ref{bell_optimization} shows the optimization trajectory of Bell state creation. The model quickly starts to converge, and after some time we obtained Bell state with a fidelity close to 80\%. Since only half of the excitation is being transmitted in the process, the transmission is less likely to be lost. We are able to obtain bell state creation with a fidelity higher than single photon transfer. During the optimization, we clipped the value of density matrix to a maximum of 0.5 for the calculation of fidelity. Without doing so, we found our numerical optimization results bias towards a higher excited population ($>$ 0.5) of the sender qubit, where ideally one would expect the excited population to be 0.5. This artifact is likely due to the inner product definition of the fidelity, where $>$ 0.5 excited population actually increases part of the inner products. We also took the absolute value of the resulting density matrix. This process optimized the Bell state up to a local qubit phase. To recover the target Bell state, we used the transfer parameters obtained from the optimizer and applied local phase advancement on one of the qubits. We repeated the Bell state creation experiment for 10+ times to obtain a statistics on the error of the Bell state fidelity. The resulted Bell state fidelity with this procedure was 79.3\%  $\pm$ 0.3\%. The online optimization with Gaussian process works reasonably well even we started with random initial parameters for the two square pulses. For arbitrarily shaped pulses, the high-dimensionality would necessarily require one to employ a model-based offline quantum optimal control \cite{Leung2017SpeedupUnits} to facilitate the optimization process.

\end{document}